\documentclass[twocolumn]{aastex7}
\usepackage{rotating}
\usepackage{comment}
\usepackage[maxfloats=256]{morefloats}
\received{}
\revised{}
\accepted{}
\shorttitle{The chemical yields of stars in the range 9-15 $\rm M_\odot$}
\shortauthors{Limongi et al.}
\graphicspath{{./}{figures/}}

\newcommand\nuk[2]{$\rm ^{\rm #2} #1$}

\begin{document}
\title{The chemical yields of stars in the mass range 9-15 $\rm M_\odot$}
\correspondingauthor{Marco Limongi}

\email{marco.limongi@inaf.it}

\author[0000-0002-3164-9131]{Marco Limongi}
\affiliation{Istituto Nazionale di Astrofisica - Osservatorio Astronomico di Roma, Via Frascati 33, I-00040, Monteporzio Catone, Italy}
\affiliation{Kavli Institute for the Physics and Mathematics of the Universe (WPI), The University of Tokyo Institutes for Advanced Study, The University of Tokyo, Kashiwa, Chiba 277-8583, Japan}
\affiliation{INFN. Sezione di Perugia, via A. Pascoli s/n, I-06125 Perugia, Italy}
\email{marco.limongi@inaf.it}

\author[0000-0003-0390-8770]{Lorenzo Roberti}
\affiliation{Istituto Nazionale di Fisica Nucleare - Laboratori Nazionali del Sud, Via Santa Sofia 62, Catania, I-95123, Italy}
\affiliation{Konkoly Observatory, Research Centre for Astronomy and Earth Sciences, E\"otv\"os Lor\'and Research Network (ELKH), Konkoly Thege Mikl\'{o}s \'{u}t 15-17, H-1121 Budapest, Hungary}
\affiliation{CSFK, MTA Centre of Excellence, Budapest, Konkoly Thege Mikl\'{o}s \'{u}t 15-17, H-1121, Hungary}
\affiliation{Istituto Nazionale di Astrofisica - Osservatorio Astronomico di Roma, Via Frascati 33, I-00040, Monteporzio Catone, Italy}
\email{lorenzo.roberti@inaf.it}

\author[0009-0005-0324-0637]{Agnese Falla}
\affiliation{Dipartimento di Fisica, Sapienza Università di Roma, P.le A. Moro 5, Roma 00185, Italy}
\affiliation{Istituto Nazionale di Astrofisica - Osservatorio Astronomico di Roma, Via Frascati 33, I-00040, Monteporzio Catone, Italy}
\email{agnese.falla@uniroma1.it}

\author[0000-0002-3164-9131]{Alessandro Chieffi}
\affiliation{Istituto Nazionale di Astrofisica - Istituto di Astrofisica e Planetologia Spaziali, Via Fosso del Cavaliere 100, I-00133, Roma, Italy}
\affiliation{Monash Centre for Astrophysics (MoCA), School of Mathematical Sciences, Monash University, Victoria 3800, Australia}
\affiliation{INFN. Sezione di Perugia, via A. Pascoli s/n, I-06125 Perugia, Italy}
\email{alessandro.chieffi@inaf.it}

\author[0000-0001-9553-0685]{Ken'ichi Nomoto}
\affiliation{Kavli Institute for the Physics and Mathematics of the Universe (WPI), The University of Tokyo Institutes for Advanced Study, The University of Tokyo, Kashiwa, Chiba 277-8583, Japan}
\email{nomoto@astron.s.u-tokyo.ac.jp}

\begin{abstract}
In \cite{lrc24} we presented and discussed the main evolutionary properties and final fate of stars in the mass range $\rm 7-15~M_\odot$. The evolutions of those models were computed by means of a medium size nuclear network that guaranteed a proper calculation of the nuclear energy generation and hence a good modeling of the physical evolution of these stars. In the present paper, we extend this study by computing the detailed chemical yields of stars in the mass range $\rm 9–15~M_\odot$, i.e., those stars that explode as core collapse supernovae (CCSNe). The explosive nucleosynthesis is then computed in the framework of the thermal bomb induced explosion by means of the HYPERION code \citep{lc20a}. We find that: (1) the yields of the intermediate mass elements (i.e., O to P) show a steep decrease as the inital mass decreases; (2) the yields of s-weak component, i.e., those produced by the slow neutron captures from Ga to to the first neutron closure shell, decrease almost linearly as a function of the initial mass with respect to the ones produced by the more massive stars; (3) the global contribution of the stars in the mass range $\rm 9.22-13~M_\odot$ to the yields of a generation of massive stars averaged over a standard initial mass function is negligible for essentially all the isotopes. In spite of this, however, the models of stars in this mass range can be fundamental to interpret the observations of specific supernovae. 
\end{abstract}

\section{Introduction}\label{sec:intro}
According to a standard Salpeter Initial Mass Function (IMF), stars in the mass range $7-12~M_\odot$ constitute $\sim 50\%$ of all the stars with mass greater than $\rm 7~M_\odot$. Therefore, a proper understanding of their evolution and final fate is necessary for a proper modeling of the physical and chemical evolution of the galaxies and, more in general, of the Universe. In \citet[][PaperI]{lrc24} we presented and discussed detailed evolutions and final fate of non-rotating stars in the mass range $\rm 7-15~M_\odot$ with initial solar metallicity and concluded that: (1) stars with initial mass $\rm M\leq7.00~M_\odot$ develop a degenerate CO core, evolve through the thermally pulsing asymptotic giant branch phase (AGB), and eventually end their evolution as a CO White Dwarf (CO-WD); (2) stars with initial mass in the range $\rm 7.50\leq M/M_\odot\leq 9.22$ develop a degenerate ONeMg core and evolve through the super-AGB phase (SAGB): within this mass interval stars with initial mass in the range $\rm 7.50\leq M/M_\odot\leq 8.00$ end their lives as CONe WDs, (Chieffi et al., 2025, submitted), sometimes called hybrid WD or ONeMg WDs \citep{nomoto84}, while stars with initial mass in the range $\rm 8.50\leq M/M_\odot\leq 9.20$ may potentially explode as electron capture supernovae \cite[ECSN,][]{nomoto84,nomoto87}; (3) stars with initial mass $\rm M\geq 9.22~M_\odot$ evolve through all the major nuclear burning stages, form an iron core and eventually explode as core collapse supernovae (CCSN).

The mass boundaries discussed above are generally influenced by the initial conditions (initial metallicity and rotation), and by various uncertainties in stellar modeling \citep{doherty17}. Two particularly significant uncertainties are the convective core overshooting during core hydrogen burning and the mass loss. Convective core overshooting determines the relationship between the initial mass and the mass of the He and subsequent CO cores, the CO core mass being one of the key drivers of stellar evolution after core He burning. Mass loss, on the other hand, plays a crucial role in the competition between the CO core growth and the reduction of the H-rich envelope, ultimately deciding the star's fate: if the core growth prevails, it may explode as an ECSN, whereas if mass loss dominates, it will evolve into an ONeMg White Dwarf.

The models presented in Paper I were computed by means of the latest version of the FRANEC code \citep{cl13,lc18} with a medium size nuclear network. The choice of such a network was mainly due to the fact that in that paper we were mainly interested in the proper modeling of the physical evolution of these stars (and not in the detailed tracing of their chemical evolution) and hence we have chosen the smallest nuclear network that guaranteed the computation of the nuclear energy generation with a very good accuracy in all the evolutionary phases. This allowed us to properly follow the physical evolution of these stars with a great numerical resolution in a reasonable amount of time.

In this paper, we present the chemical yields of stars in the range $\rm 9.22\leq M/M_\odot\leq 15.00$, i.e., stars that explode as CCSNe. To accomplish this goal, we recomputed the pre-supernova evolution of stars with initial mass $\rm 9.22,~10,~11,~12,~13,~and~15~M_\odot$ up to the onset of the iron core collapse, with the same code and input physics adopted in Paper I but with a much more extended nuclear network in order to trace the abundance of a large number of isotopes. We also take into account the explosive nucleosynthesis by computing induced explosions by means of the HYPERION code \citep{lc20a}.

\section{Numerical Methods and Nuclear Nuclear Network}\label{sec:numerics}
All the models presented in this paper have been computed by means of the same stellar evolution code and input physics described in Paper I, therefore we refer the reader to that paper for all the details. The only difference with respect to the calculations presented in Paper I is the adoption of a much larger nuclear network.
More specifically, in Paper I we have adopted a 112 isotope network (from neutrons to $\rm ^{60}Ni$, see Table 1 in Paper I) that guaranteed the calculation of the nuclear energy generation with great accuracy. In this paper, on the contrary, we adopted the same nuclear network used in \citet[][see their Table 1]{lc18}, that includes 335 nuclear species (from neutrons to $\rm ^{209}Bi$) and $\sim 3100$ nuclear reactions.

For all the isotopes lighter than $\rm ^{98}Mo$ we consider all the possible reactions involving the capture or emission of a proton, $\alpha$-particle, neutron or photon. All the possible reactions due to the weak interactions (electron or positron captures and $\beta$-decays) are also taken into account. Also included are the triple-$\alpha$, $\rm ^{12}C+^{12}C$, $\rm ^{12}C+^{16}O$ and $\rm ^{16}O+^{16}O$ reactions. For temperatures above $T=4\cdot10^{9}~{\rm K}$ we shift to a full Nuclear Statistical Equilibrium (NSE) network, as described in \citet{cls98}. For isotopes heavier than $\rm ^{98}Mo$ we consider only the neutron captures and $\beta$-decays. As discussed in detail in \citet{lc18}, the isotopes between $\rm ^{98}Mo$ and $\rm ^{132}Xe$ and between $\rm ^{144}Nd$ and $\rm ^{202}Hg$ are assumed to be at the local equilibrium. 
The reliabilty of this assumption has been clearly demonstrated in \citet{lc18} with a one-zone calculation of a typical He burning condition performed with two different nuclear networks, i.e., the one with the present network and another one in which we added all the stable isotopes between $\rm ^{98}Mo$ and $\rm ^{132}Xe$ \citep[see Table 2 in][]{lc18}. A detailed discussion of the result of this test is addressed in \citet{lc18}.
The nuclear cross sections and the weak interaction rates are the same reported in Table 3 in \cite{lc18}.

The explosive nucleosynthesis has been computed by means of the HYPERION code \citep{lc20a}. Although the code as well as the induced explosion technique are well described in \cite{lc20a}, we summarize here the basic assumptions. The explosion is induced by firstly removing the inner 0.8 $\rm M_\odot$ of the presupernova model and at this mass coordinate we instantaneously deposit a given amount of thermal energy spread over $\rm 0.1~M_\odot$. The propagation of the shock wave that forms as a consequence of this energy injection is then followed by solving the hydrodynamical equations and the radiation transport equation in the flux-limited diffusion approximation. The same nuclear network adopted for the hydrostatic evolutions is coupled to these equations in order to properly trace the abundance of all the nuclear species considered. The energy initially injected is calibrated in order to obtain a final kinetic energy of the ejecta of $\rm \sim 10^{51}~erg$. 
Such a choice assures that all models eject the whole mass above the Fe core. The mass cut, i.e., the mass coordinate that separates the remnant from the ejecta, is then fixed "a posteriori" on the basis of the chosen strategy, for example by requiring the ejection of a given amount of \nuk{Ni}{56} \citep[e.g.,][]{lc03,nkt13} or an initial mass-remnant mass relation or an initial mass-ejected $\rm ^{56}Ni$ mass relation. In this paper we fix the mass cut by assuming the initial mass-ejected $\rm ^{56}Ni$ mass relation obtained by \cite{burrows+24} by means of full 3D hydrodynamical simulations. In particular, we assume the following ejected $\rm ^{56}Ni$ mass: $\rm 1.04\cdot 10^{-2}~M_\odot$, $\rm 1.95\cdot 10^{-2}~M_\odot$, $\rm 2.92\cdot 10^{-2}~M_\odot$, $\rm 3.55\cdot 10^{-2}~M_\odot$, $\rm 4.17\cdot 10^{-2}~M_\odot$ and $\rm 5.42\cdot 10^{-2}~M_\odot$ for the 9.22, 10, 11, 12, 13 and $\rm 15~M_\odot$ models respectively. However, as it has been done in \cite{lc18}, we provide, online and upon request, all the data necessary to compute the yields under different assumptions.


\section{The models}\label{sec:models}
We computed the presupernova evolution of stars with initial mass $9.22$, $10$, $11$, $12$, $13$, and $\rm 15~M_\odot$, from the pre-main sequence phase up to the onset of the iron core collapse. Their main evolutionary properties are reported in Table \ref{tab_main_prop}. 

The various entries have the following meaning: $\rm M_{CC}$ is the maximum size of the convective core in units of $\rm M_\odot$; $\rm t$ is the evolutionary time in units of yr; $\rm ^{12}C$ is the central carbon mass fraction (this entry is reported only for the He burning phase and refers to the value at core He depletion); $\rm M_{Fe}$, $\rm M_{SiS}$, $\rm M_{ONe}$, $\rm M_{CO}$, $\rm M_{He}$, $\rm M_{CE}$ and $\rm M_{tot}$ are the iron core mass, the SiS (O depleted) core mass , the ONe core mass, the CO core mass, the He core mass, the convective envelope mass and the total mass, respectively, in units of $\rm M_\odot$; $\rm \psi_{c}$, $\rm T_{c}~(K)$, $\rm \rho_c~(g~cm^{-3})$ and $\rm Y_{e,c}$ are the central values of the degeneracy parameter, the temperature, the density and the electron fraction; $\rm T_{ign}~(K)$, $\rm \rho_{ign}~(g~cm^{-3})$, $\rm \psi_{ign}$ and $\rm M_{ign}~(\rm M_\odot)$ are the temperature, density, degeneracy parameter and mass coordinate corresponding to an off-center nuclear ignition; $\rm M_C$ refers to the mass coordinate, in units of $\rm M_\odot$, marking the central zone where the carbon mass fraction exceeds 0.01, this quantity being relevant for those stars that form hybrid CO/ONe cores, i.e., cores mainly composed by ONe but with a central region still rich in C.

\begin{figure*}[ht!]
\epsscale{1.1}
\plotone{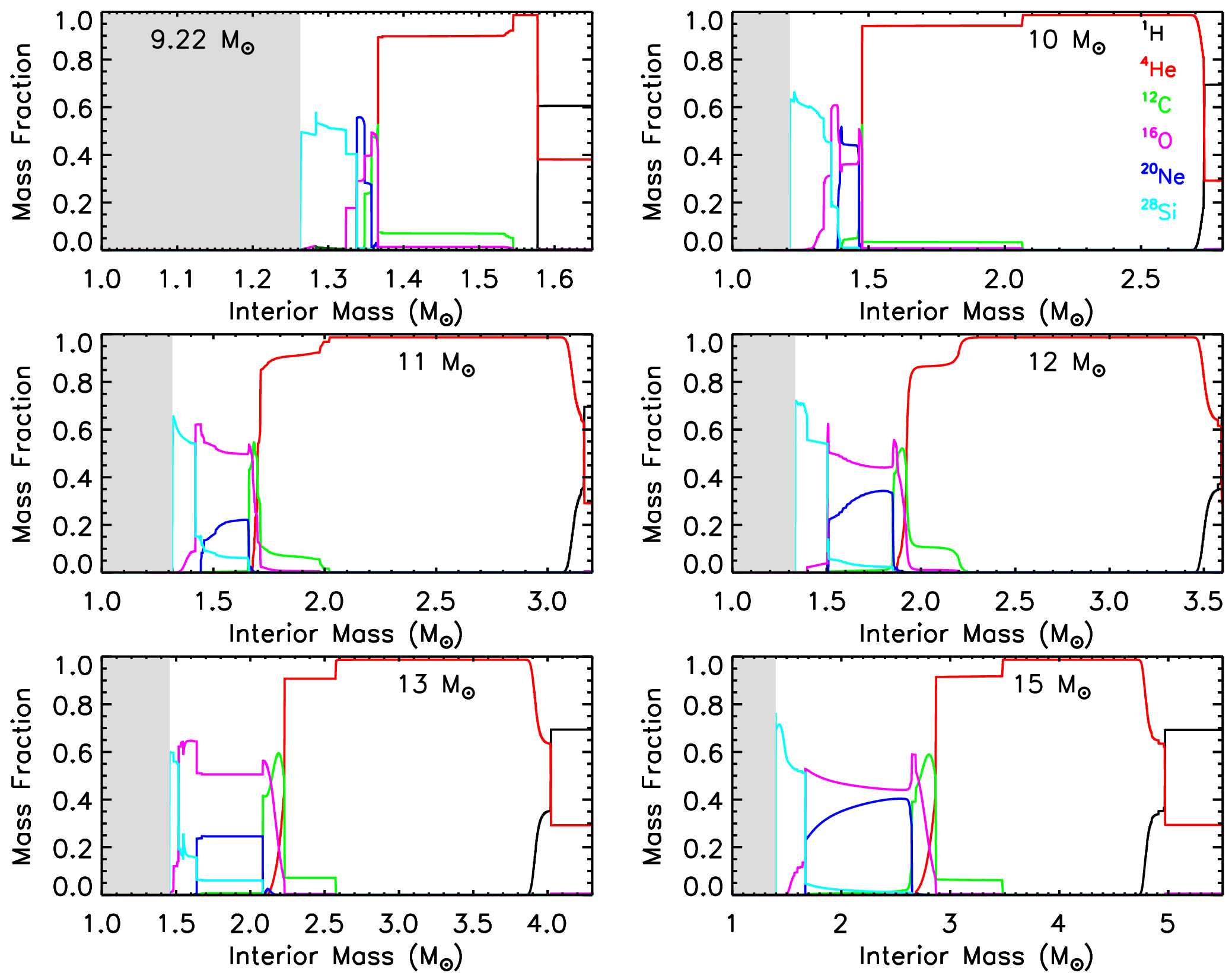}
\caption{Chemical composition of all the models at the presupernova stage.\label{fig:majorall}}
\end{figure*}

\begin{figure*}[ht!]
\epsscale{0.8}
\plotone{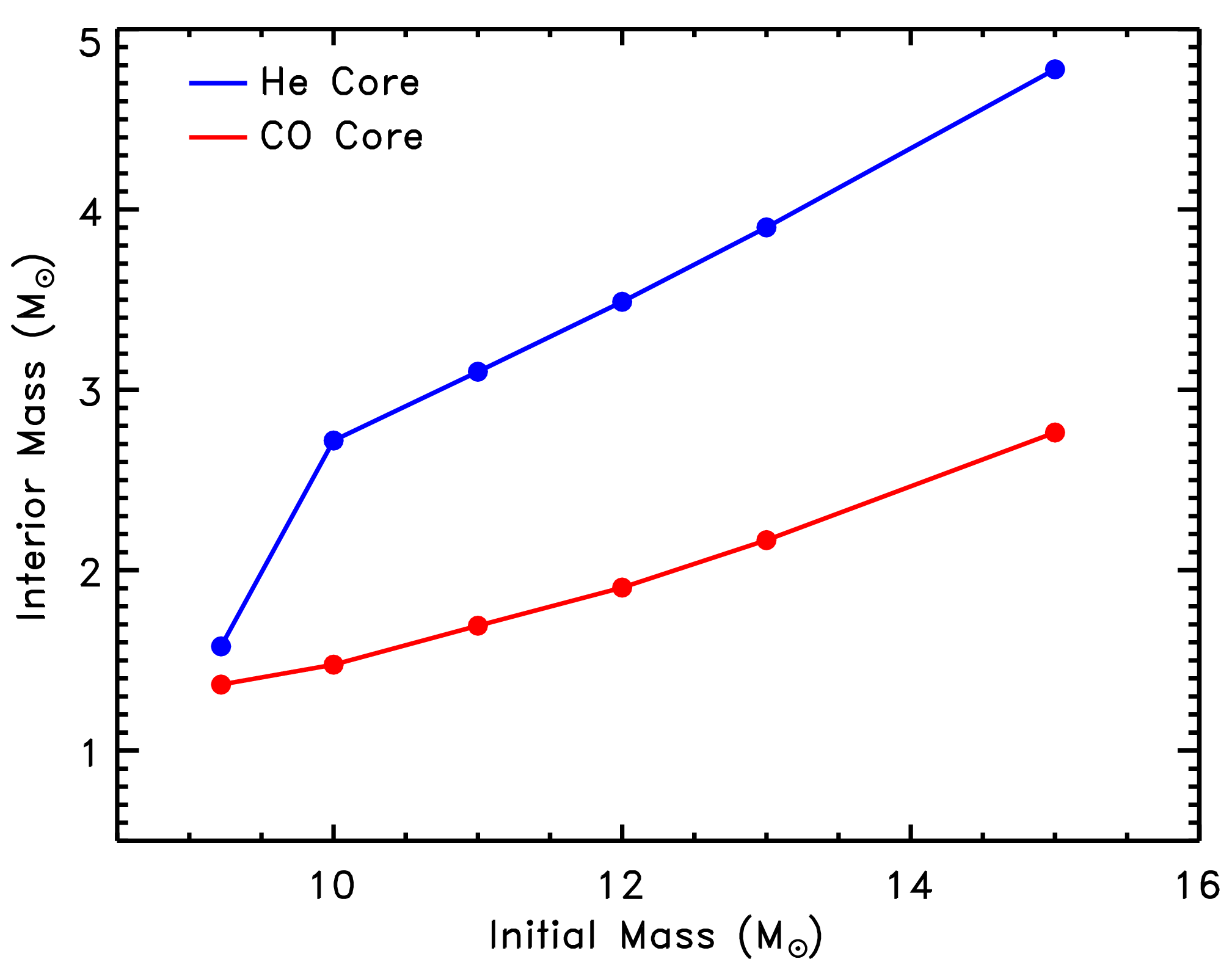}
\caption{He core mass (blue line-blue filled dots) and CO core mass (red line-red filled dots) as a function of the initial mass at the presupernova stage. \label{fig:helayer}}
\end{figure*}

Figure \ref{fig:majorall} shows the the structural profiles of the most abundant isotopes of all the models at the presupernova stage. The Figure clearly shows that the mass of the He layer, which is confined between the CO core mass and the He core mass, decreases substantially as the initial mass decreases (see also Figure \ref{fig:helayer}). Such an occurrence is the consequence of the second dredge up that occurs only in the less massive model of this grid, i.e., the $\rm 9.22~M_\odot$, reminiscence of the formation of a partial electron degenerate core, see Figure 3 in \citet{lrc24}. 

\begin{figure*}[ht!]
\epsscale{0.8}
\plotone{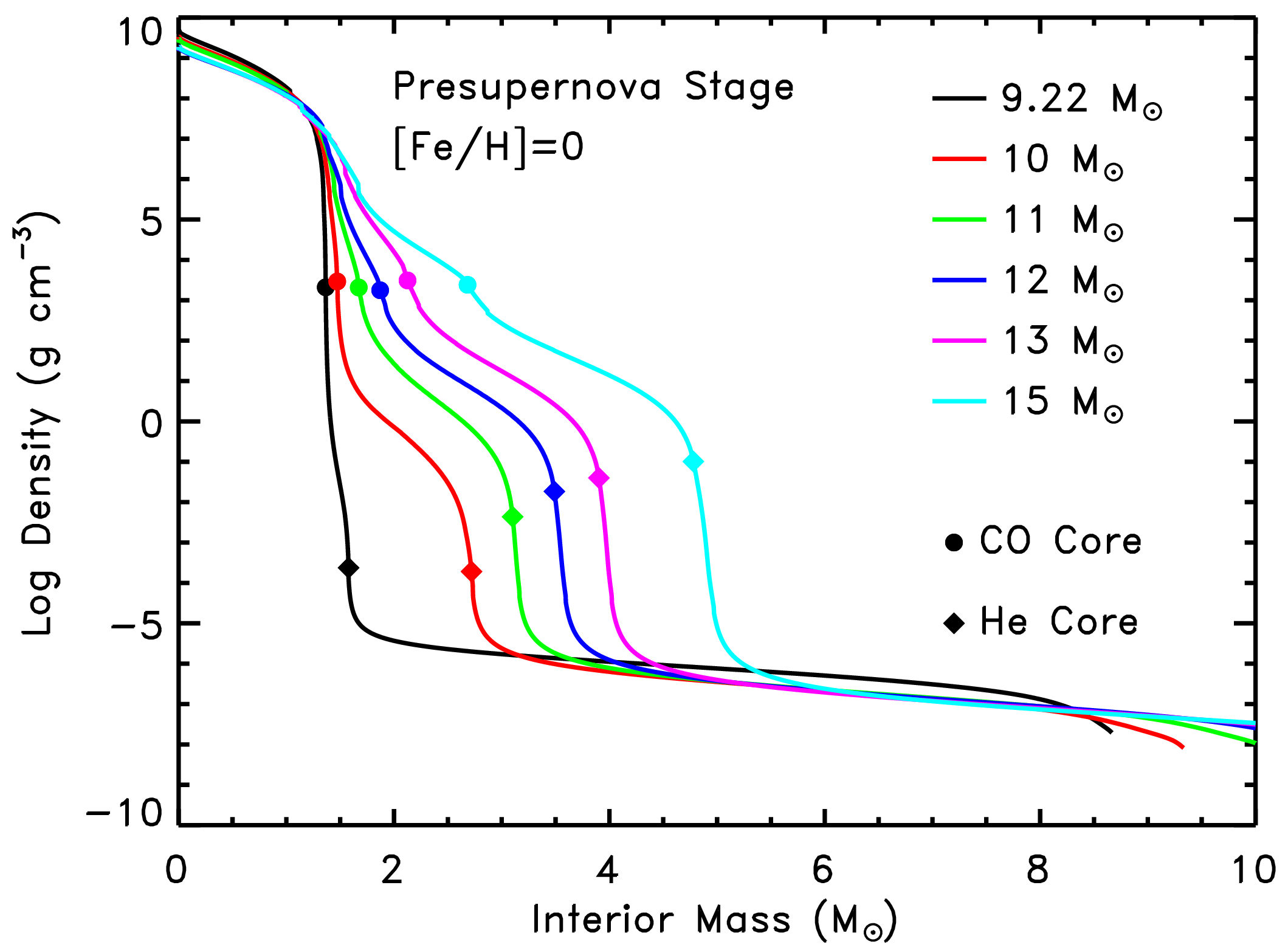}
\caption{Density profiles of all the models at the presupernova stage. The filled dots and the filled rombs mark the CO core and the He core respectively. \label{fig:tempdens}}
\end{figure*}

The strong reduction of the He core mass that occurs in the smallest mass of our grid as a consequence of the second dredge up, creates a sharp discontinuity in the general trend of the final density gradient versus the initial mass. Figure \ref{fig:tempdens} 
shows, in fact, that the progressive steepening of the density gradient as the initial mass decreases, turns almost vertical in the transition between the 10 and the $\rm 9.22~M_\odot$. Such a sharp change in the density profile has important consequences on the development of the shock wave \citep{boccioli+23} and on the yields produced by the most internal explosive burning (see below).

\section{Chemical Yields}\label{sec:yields}

\begin{figure*}[ht!]
\epsscale{1.15}
\plotone{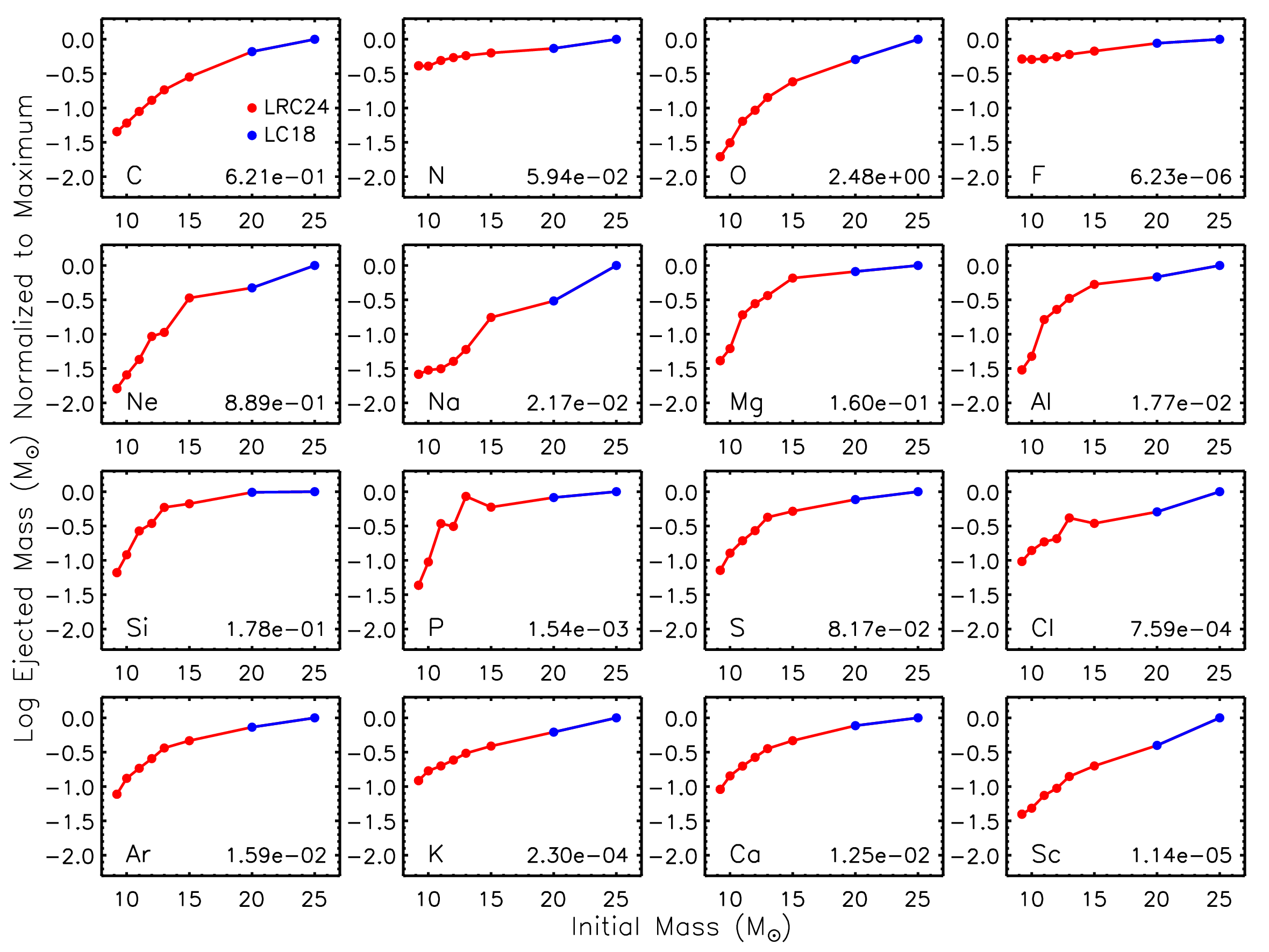}
\caption{Abundances of the various elements in the ejecta as a function of the initial mass, after the full decay of all the unstable nuclear species. To improve the readability of the figure, for each element we have normalized the abundances obtained for all the stars to the maximum one (reported in solar masses in the lower corner of each panel). The red lines-red filled dots refer the results obtained in this study while the blue lines-blue filled dots to the values published by \cite{lc18}. \label{fig:yieldsNi007_1}}
\end{figure*}

\begin{figure*}[ht!]
\epsscale{1.15}
\plotone{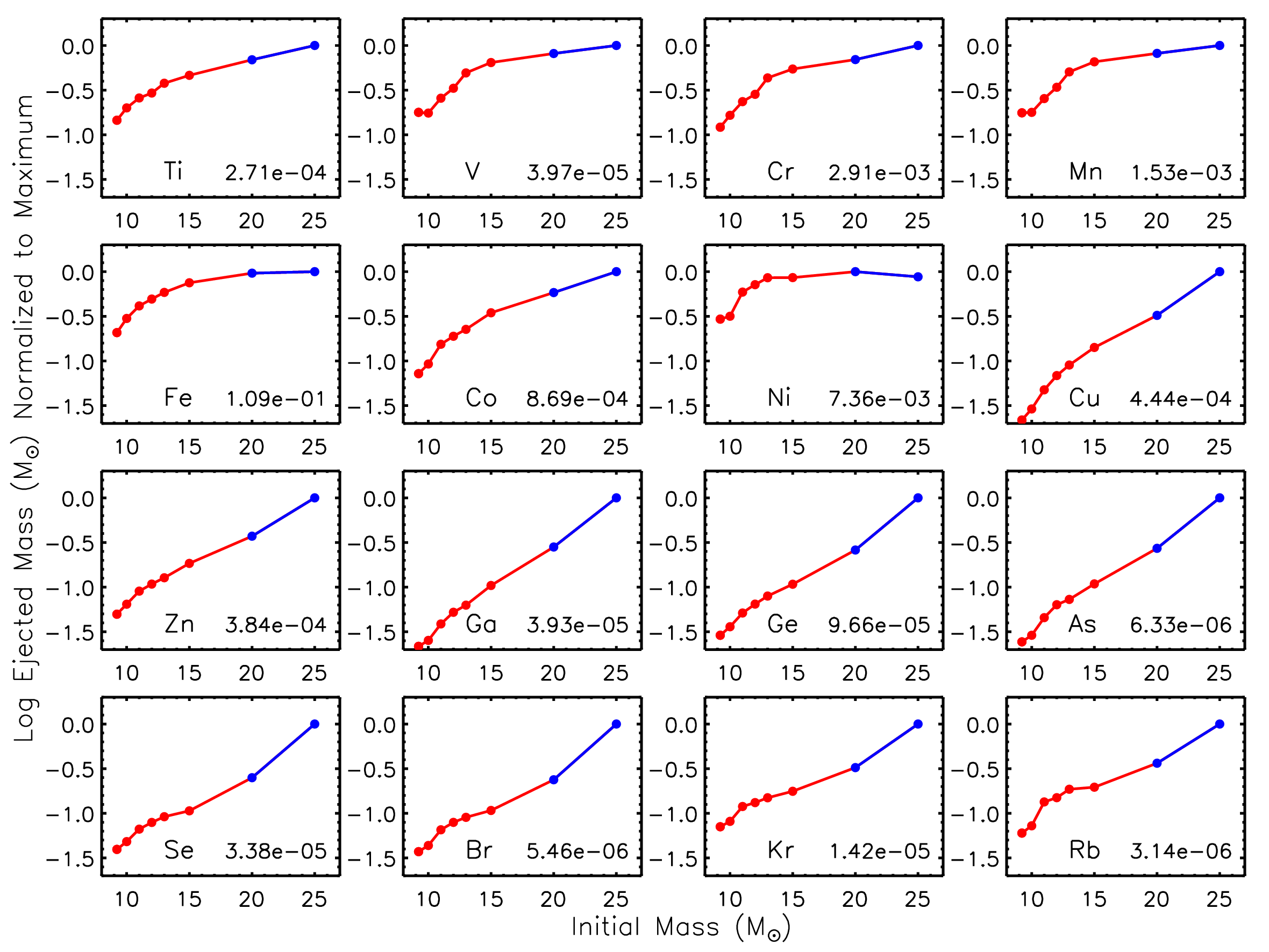}
\caption{Same as Figure \ref{fig:yieldsNi007_1} \label{fig:yieldsNi007_2}}
\end{figure*}

\begin{figure*}[ht!]
\epsscale{0.8}
\plotone{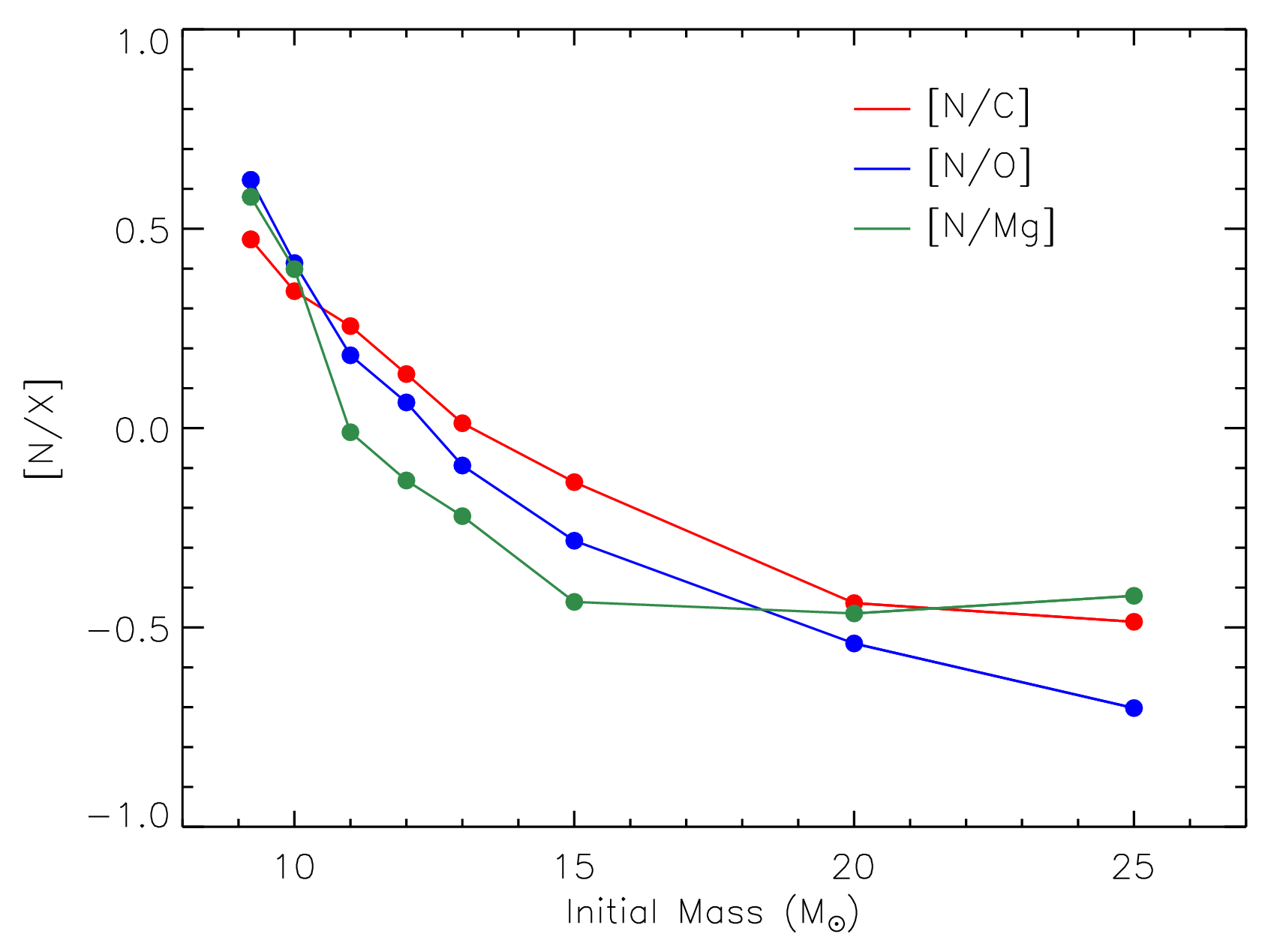}
\caption{Selected element abundance ratios in the ejecta of stars with initial mass in the range $\rm 9.22-15~M_\odot$. \label{fig:NsuO}}
\end{figure*}

Figures \ref{fig:yieldsNi007_1} and \ref{fig:yieldsNi007_2} show the yields of some selected elements as a function of the initial mass obtained by fixing the mass cut as described in section \ref{sec:numerics}. The red filled circles in Figures \ref{fig:yieldsNi007_1} and \ref{fig:yieldsNi007_2} refer to the present calculations while the blue filled ones to the values published in \cite{lc18}. Looking at the elements C-Sc, the first thing worth noting from Figure \ref{fig:yieldsNi007_1} is that the yields of all the alpha elements (C, O, Ne, Mg, Si, S, Ar, and Ca) decrease significantly as the initial mass decreases below $\rm \sim 15~M_\odot$, even by more than an order of magnitude. On the contrary, some of the odd-Z elements behave like the alpha elements (as, e.g, Na, Al and P), while some others show a much smoother decrease as the initial mass decreases (as, e.g., N, F, K, and Sc). A first conclusion of this analysis is that a linear extrapolation of the ejected masses (in a logarithm) of most of these elements down to the lower mass stars overestimate, even substantially, the total yields. Moreover, a different behavior found between all even and some odd-Z elements could be used as a proxy for a supernova progenitor in this mass range. As an example we show in Figure \ref{fig:NsuO} how some observed element abundance ratios in the spectra of a core collapse supernova can be used to infer a low mass progenitor star.

The elements heavier than Zn, i.e., those produced by the slow neutron captures and shown in Figure \ref{fig:yieldsNi007_2}, appear to almost linearly (in logarithm) decrease with the initial mass. With the choice of the mass cut discussed in section \ref{sec:numerics}, the iron peak elements, i.e., those between Ti and Zn, decrease in general with the initial mass. 

\begin{figure*}
\epsscale{0.9}
\plotone{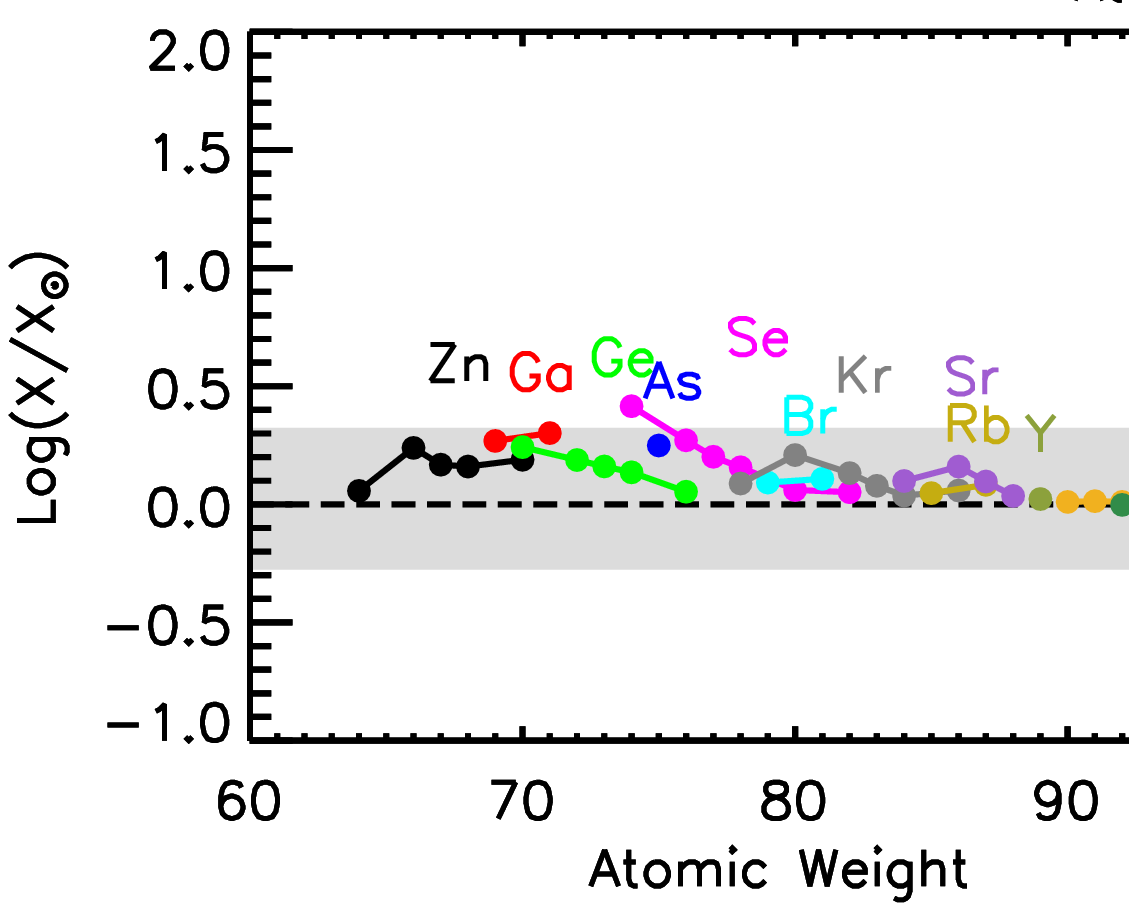}
\caption{Chemical composition of the ejecta of the $\rm 9.22~M_\odot$ star compared to the solar composition. The grey band correspond to a variation by a factor of 2 with respect to  $\rm Log(^{16}O/^{16}O_\odot)$. \label{fig:isoyieldsNi007_922} }
\end{figure*}

\begin{figure*}
\epsscale{0.9}
\plotone{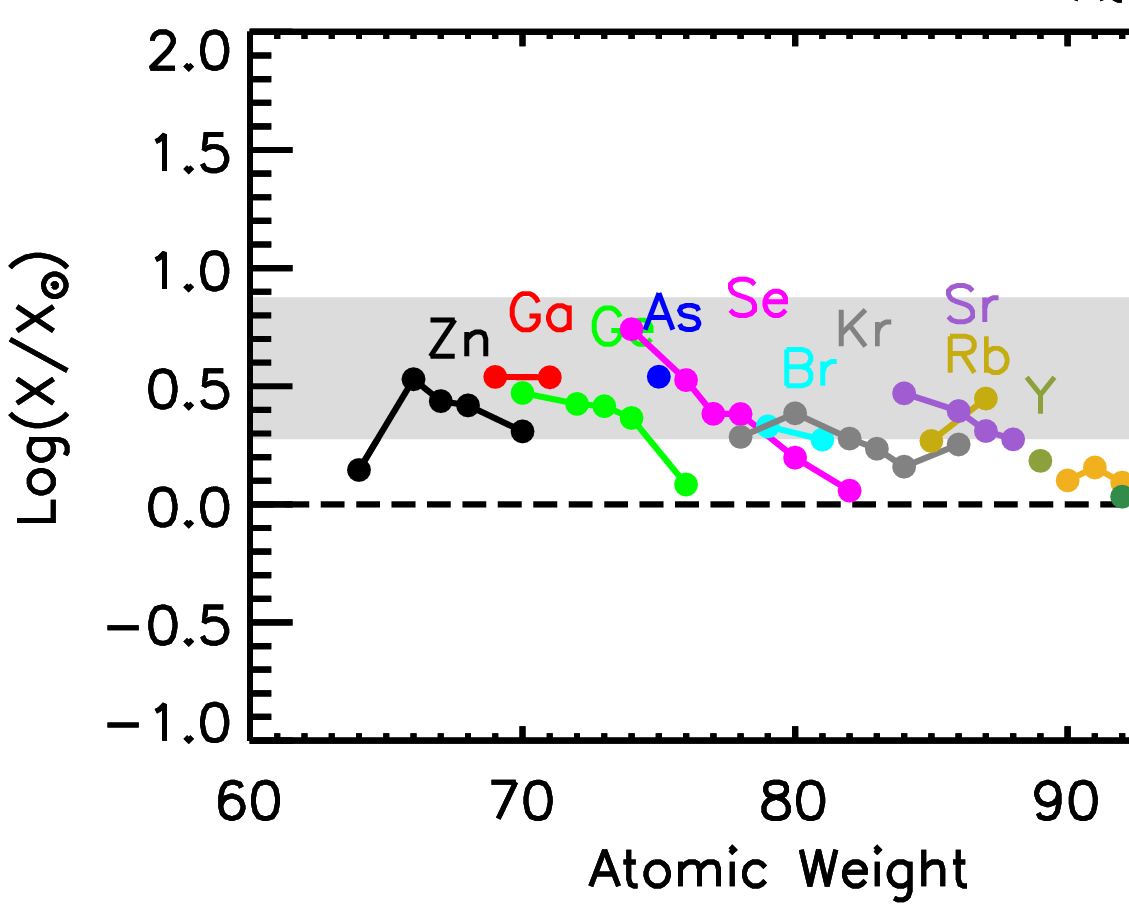}
\caption{Same as Figure \ref{fig:isoyieldsNi007_922} but for the $\rm 12~M_\odot$ star.\label{fig:isoyieldsNi007_1200}}
\end{figure*}

\begin{figure*}
\epsscale{0.9}
\plotone{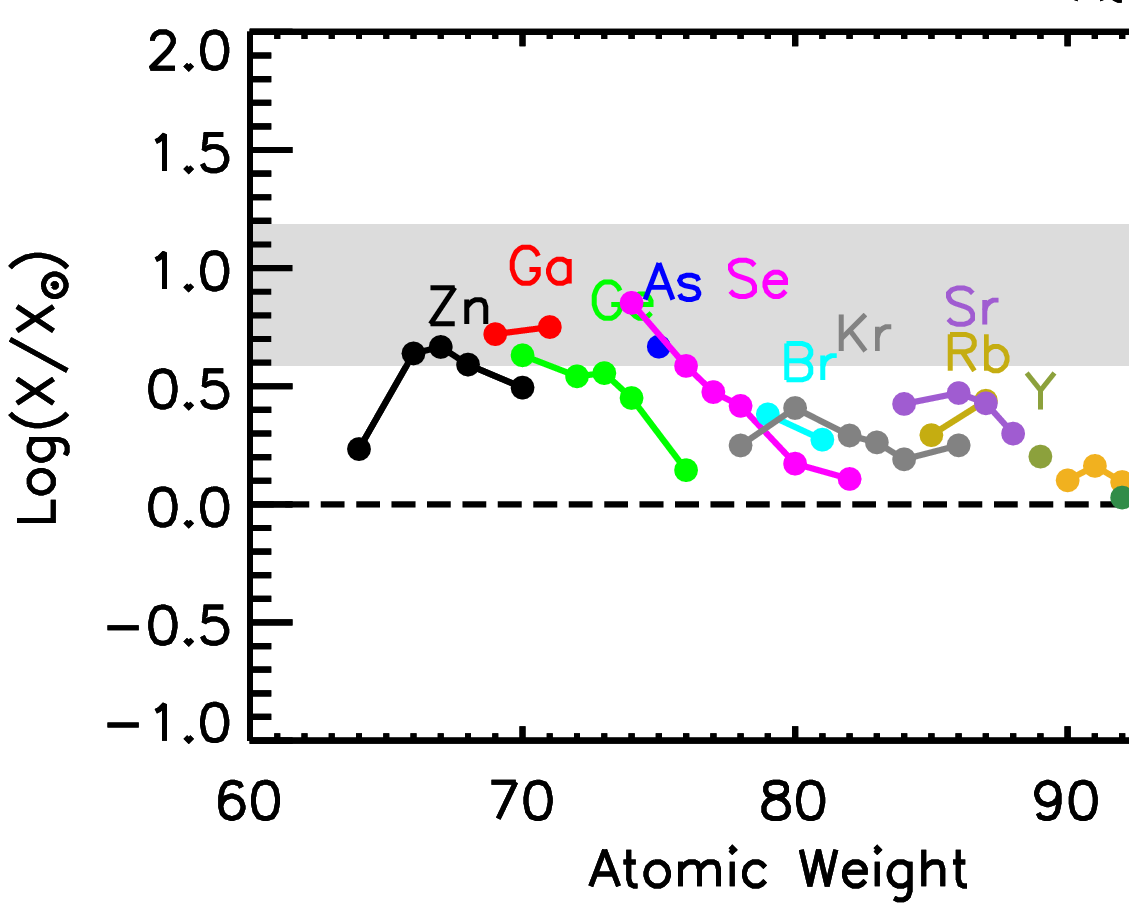}
\caption{Same as Figure \ref{fig:isoyieldsNi007_922} but for the $\rm 15~M_\odot$ star.\label{fig:isoyieldsNi007_1500}}
\end{figure*}

Figures 
\ref{fig:isoyieldsNi007_922},
\ref{fig:isoyieldsNi007_1200} and
\ref{fig:isoyieldsNi007_1500} show a comparison between the isotopic distribution in the ejecta of selected models with the solar composition. The gray band in all the figures represents a factor of 2 variation with respect to $\rm Log(^{16}O/^{16}O_\odot)$. This allows us to evaluate how much the distribution of the ejecta deviates from a scaled solar one, taken $\rm ^{16}O$ as a reference isotope since it is the most abundant one after H and He, and also it is one of the main products of massive stars.

A clear trend with the mass is not evident from all these figures. In general, the production of most of the isotopes from O to Ca increases with increasing the mass of the star and that all these elements roughly keep a scaled solar distribution with some exceptions. Moreover, the ratio between the average overproduction of the isotopes from O to S and the one of the isotopes from Cl to Ca increases with the progenitor mass.

The production of the neutron capture elements Ga-Zr of the weak-s component also increases with the mass of the star with a following clear trend: (1) the [weak-s/O] ratios scale inversely with the initial mass since the production of $\rm ^{16}O$ strongly increases as the initial mass increases; (2) for each one of these elements the more neutron rich isotopes are in general underproduced with respect to the less neutron rich ones because of the low neutron density achieved during the s-process nucleosynthesis (see, e.,g. the isotopes of Ga and Se in Figures \ref{fig:isoyieldsNi007_922}, \ref{fig:isoyieldsNi007_1200} and \ref{fig:isoyieldsNi007_1500}). Elements heavier than Zr are not produced at all.

The production of the iron peak isotopes naturally depends on the location of the mass cut (see also Figure \ref{fig:yieldsNi007_2}). With the choice for the mass cut discussed above, a sizeble overproduction of all these isotopes is found relative to $\rm ^{16}O$ for the lower mass models, especially for the Ni isotopes. Such an overproduction reduces as the initial mass of the star increases. This is the natural consequence of the reduced He core (where O is produced) due to the occurrence of the dredge up (see above) and the internal location of the mass cut in the lower mass stars.

\begin{figure*}[ht!]
\epsscale{0.9}
\plotone{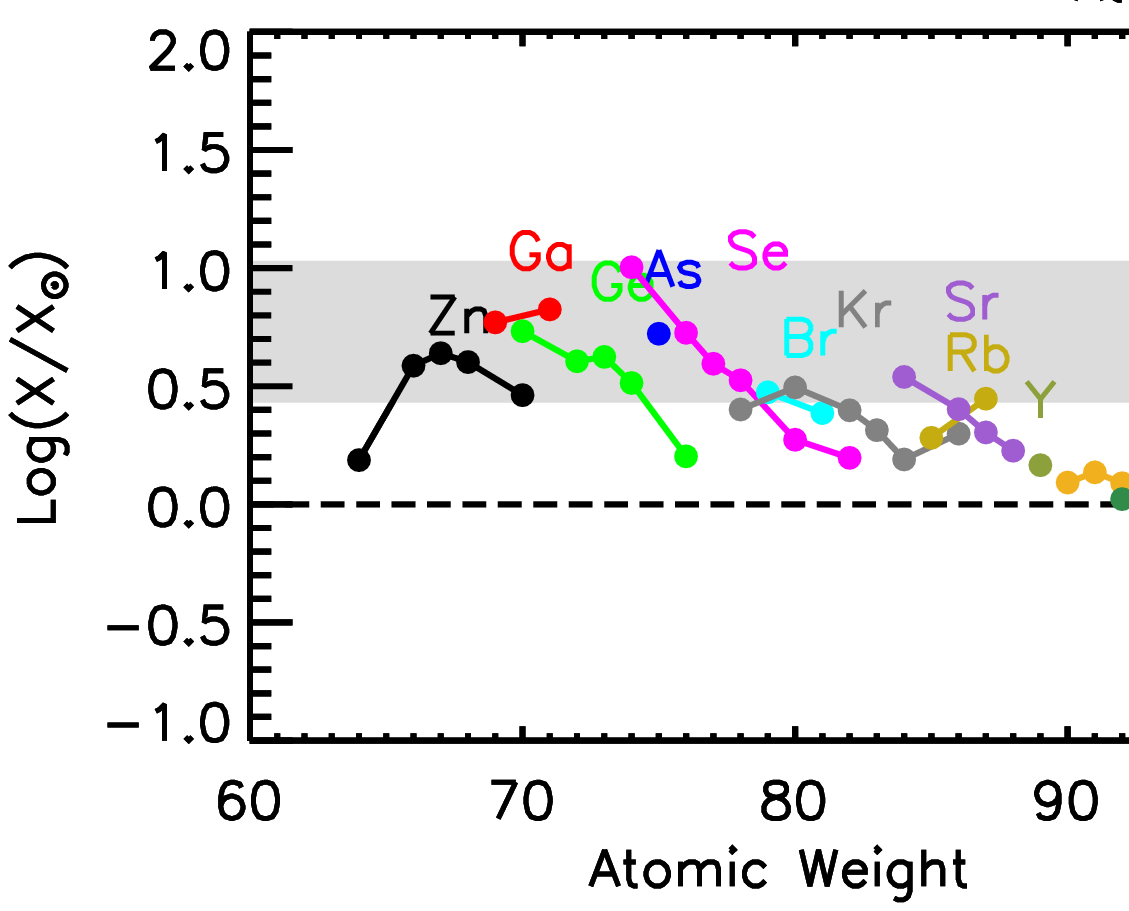}
\caption{Chemical composition of the ejecta of a generation of stars in the mass range $\rm9.22-120~M_\odot$ averaged over a Salpeter IMF compared to the solar composition. The grey band correspond to a variation by a factor of 2 with respect to  $\rm Log(^{16}O/^{16}O_\odot)$. For stars more massive than $\rm 15~M_\odot$ we have used the yields provided by \cite{lc18} under the assumption that all the stars with initial mass lower than $\rm 30~M_\odot$ eject $\rm 0.07~M_\odot$ of $\rm ^{56}Ni$ while all the more massive ones fully collapse to a black hole and hence contribute to the chemical enrichment only through the mass ejected by stellar wind.\label{fig:plot_iso_int}}
\end{figure*}

Although we have discussed above the general trends of the isotopic yields as a function of the initial mass, we report the isotopic yields averaged over a Salpeter IMF in the mass range $\rm 9.22-120~M_\odot$ in Figure \ref{fig:plot_iso_int} for sake of completeness. We find an average overproduction of the majority of the isotopes of a factor of $\rm \sim0.7~dex$ with respect to the solar values, with some exceptions like, e.g., isotopes of Cl, Ar, Ca and Ti. The isotopes $\rm ^{13}C$, $\rm ^{15}N$, $\rm ^{17}O$ and $\rm ^{19}F$ are substantially underproduced with respect to $\rm ^{16}O$, as it is expected. $\rm ^{15}N$ and $\rm ^{19}F$ are underproduced even with respect to their solar value. The neutron captures isotopes Ga-Sr are generally co-produced with $\rm ^{16}O$ with the more neutron rich ones progressively underproduced. No production is found for the isotopes of the main-s component, i.e., isotopes heavier than Zr.
The iron peak isotopes (V-Zn) are co-produced with $\rm ^{16}O$. Since most of these isotopes should be produced by other sources (like, e.g. SNIa) this means that a substantially lower amount of $\rm ^{56}Ni$ should be ejected by these stars. 

\begin{figure*}[ht!]
\epsscale{0.9}
\plotone{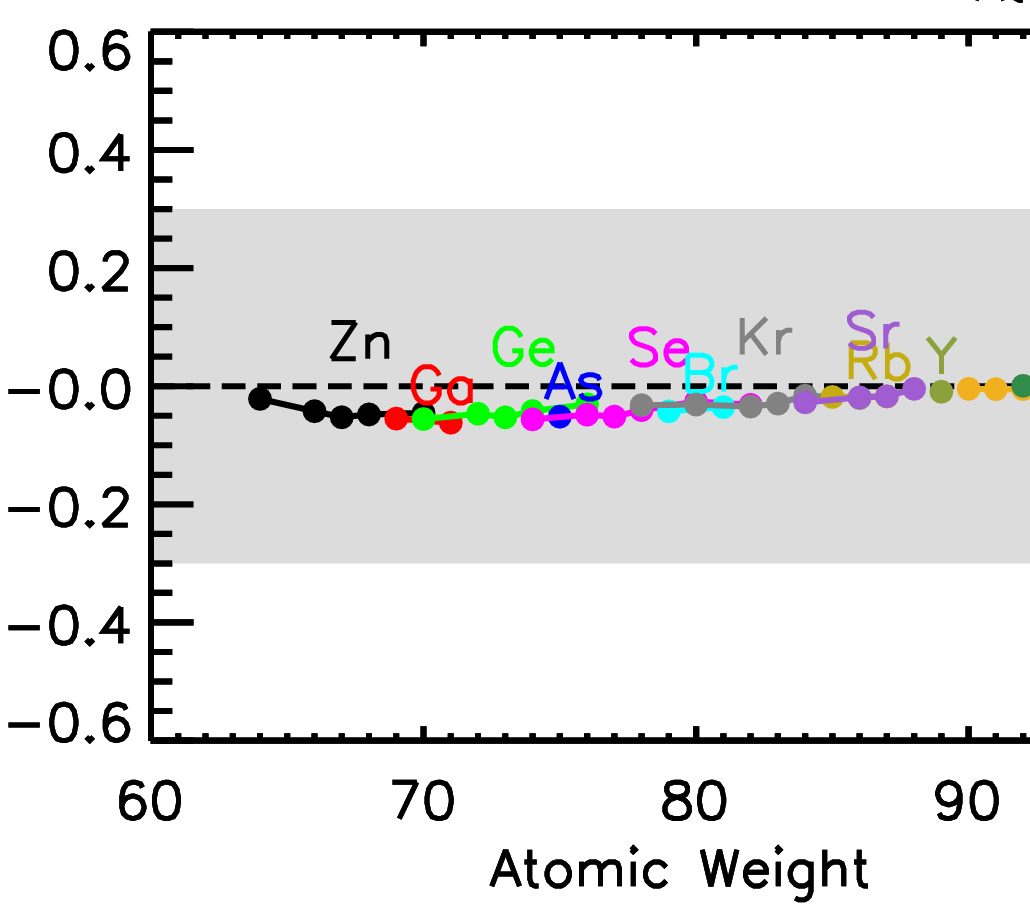}
\caption{Comparison between the chemical composition of the ejecta of a generation of massive stars averaged over a Salpeter IMF in the mass interval $\rm 9.22-120~M_\odot$ ($\rm <Yield>_{9.22-120}$) and the one averaged over the same IMF but in the mass range $\rm 13-120~M_\odot$ ($\rm <Yield>_{13-120}$). The grey band correspond to a variation of a factor of 2.\label{fig:comparexxsun_int}}
\end{figure*}

Finally, in order to study the contribution of the lower mass core collapse supernovae to the chemical enrichment due to a generation of massive stars, we have computed the element yields averaged over an initial mass function in the mass interval $\rm 9.22-120~M_\odot$ ($\rm <Yield>_{9.22-120}$) and compared these with the same calculation but this time integrating between $\rm 13-120~M_\odot$ ($\rm <Yield>_{13-120}$). Figure \ref{fig:comparexxsun_int} shows that the contribution of the stars in the mass range $\rm 9.22-13~M_\odot$ is negligible for all the isotopes at least within a factor of $\sim 2$, that we assume as a reasonable range uncertainty associated to our numerical simulations \citep[see also][]{rauscher+2002}. 

As a final comment, let us note that a comparison of the yields for the 13 and $\rm 15~M_\odot$ models with their corresponding values provided in \cite{lc18} shows not negligible differences for some elements, in particular those produced by the shell C burning, i.e., Ne, Mg, and Al. This is the consequence of the different \nuk{C}{12} mass fraction present in the core at core He exhaustion between these models. 
In particular, we obtain now for this quantity 0.412 and 0.390 compared to 0.381 and 0.343 of \cite{lc18} for the 13 and the $\rm 15~M_\odot$ models, respectively. Such a change is the result of the different strategy adopted in determining the border of the He convective core in the present models. Stars with initial mass $\rm M\lesssim 15~M_\odot$ develop a runaway of the convective core towards the end of the core He burning, i.e., the so called Breathing Pulses \citep{cast85}. Such a phenomenon brings fresh He within the convective core whose consequence is that of prolonging the conversion of $\rm ^{12}C$ into $\rm ^{16}O$. Unfortunately, at present it is not possible to determine, on the basis of first principles, if this runaway occurs or not in real stars. However, an analysis of the number ratio $\rm R_2$ between the stars on the AGB and those on the Horizontal Branch (ratio sensitive to the presence/absence of the Breathing Pulses) in a great number of galactic globular clusters let \cite{co16} to conclude that this runaway should not occur at least in low mass stars. In the present set of models we decided to follow \cite{co16} suggestion and therefore we inhibited the possible growth of this runaway, obtaining therefore a higher amount of $\rm ^{12}C$ in the He exhausted core.

\section{Light curves}
As discussed in \cite{lc20a}, the HYPERION code is able to compute the bolometric light curve associated to a given explosion. While the explosive nucleosynthesis mildly depends on the explosion energy, provided that it is not too high, i.e., of the order of $\rm 10^{52}~erg$ or more \citep{lc03,tnh96}, the light curve is significantly affected by the kinetic energy of the ejecta at the infinity as well as on the amount and degree of mixing of the ejected $\rm ^{56}Ni$ \cite[see e.g.][for a detailed discussion on this point]{lc20a}.

\begin{figure*}[ht!]
\epsscale{0.9}
\plotone{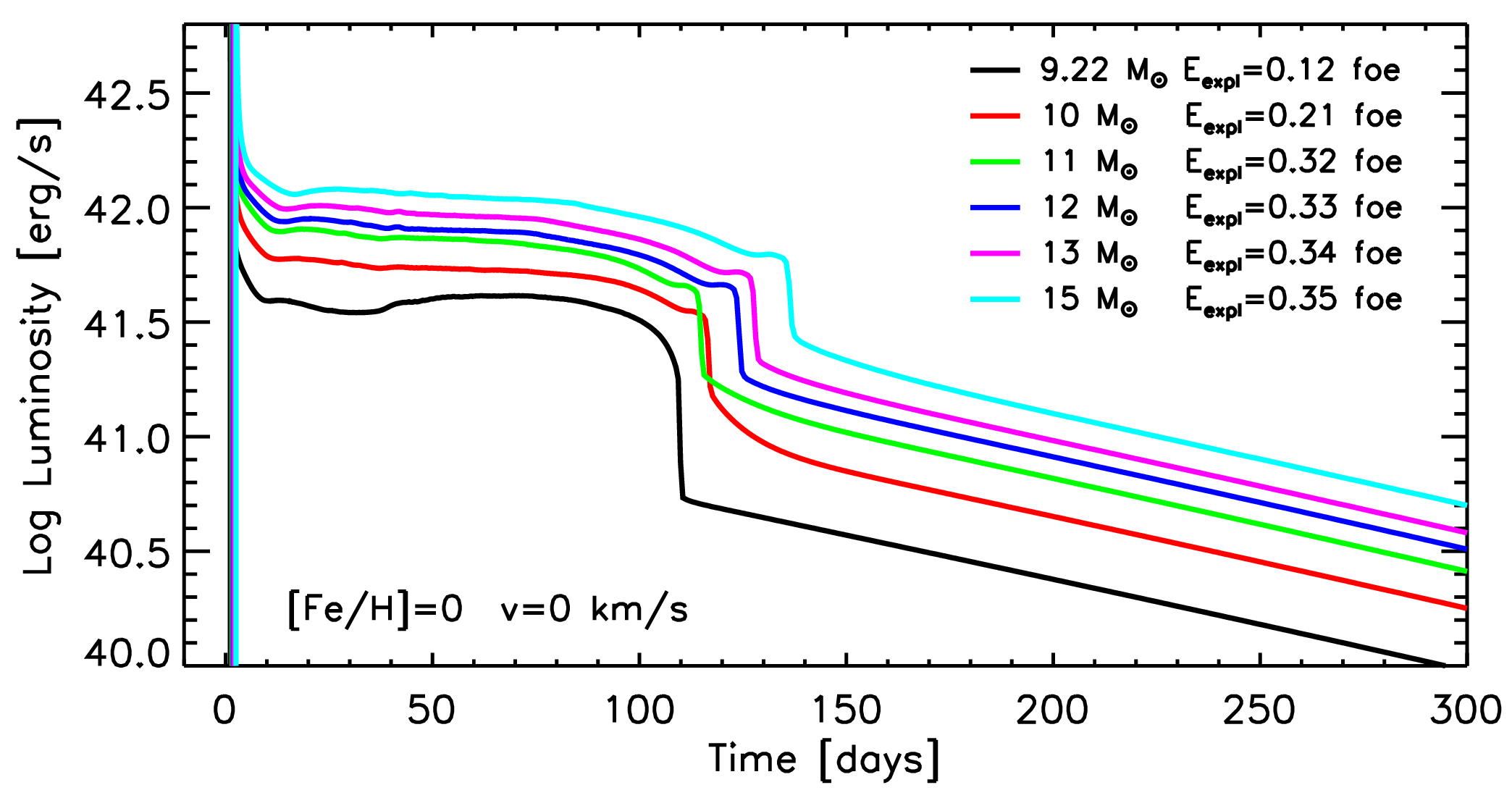}
\caption{Bolometric light curves computed assuming the initial energy-initial mass and ejected  $\rm ^{56}Ni$-initial mass relations reported in Table \ref{tab_expl_prop}. In all these calculations we assume that the $\rm ^{56}Ni$ synthesized during the explosion is mixed from the inner edge of the exploding mantle to about half of the H-rich envelope \cite{lc20a}.\label{fig:multilc}}
\end{figure*}

For this reason, in order to predict the bolometric optical display of a typical explosion of a core collapse supernova with a progenitor in this range of masses, we computed explosions and light curves assuming the initial mass-explosion energy 
relation obtained by \cite{burrows+24} 
and reported in Table \ref{tab_expl_prop}. The initial mass-ejected $\rm ^{56}Ni$ relation adopted is also the one provided by \cite{burrows+24} and it is the same adopted for the calculation of the yields shown in Figures \ref{fig:yieldsNi007_1} and \ref{fig:yieldsNi007_2} (see section \ref{sec:yields}). As discussed in \cite{lc20a}, we assume in all these calculations that the $\rm ^{56}Ni$ synthesized during the explosion is mixed from the inner edge of the exploding mantle to about half of the H-rich envelope. This mixing produces a contribution of the $\gamma$-rays to the luminosity at early times and this, in turn, implies a flatter plateau and a rapid decline of the luminosity during the transition phase from the plateau to the radioactive tail, as it is generally observed \citep[see][for a more detailed discussion on this point]{lc20a}.

Figure \ref{fig:multilc} shows the light curves of all the models computed under the assumptions discussed above. In this framework we expect that the luminosity of the plateau at 30 days after the explosion ranges between ${\rm Log}~{(L/L_\odot)\sim 41.5 }$ and ${(L/L_\odot)\sim 42.1 }$. This quantity depends mainly on the explosion energy, as it has been discussed in \cite{lc20a}. The length of the plateau, that depends on the explosion energy, the mass of the H-rich envelope and the amount and degree of mixing of the $\rm ^{56}Ni$, ranges between 110 and 140 days. The radioactive tail scales with the amount of $\rm ^{56}Ni$ ejected, as expected. Figure \ref{fig:multilc} clearly shows that the two lower mass models, i.e., the 9.22 and $\rm 10~M_\odot$, have a different behavior compared to the more massive stars. The reason is that the explosion energy does not vary smoothly with the initial mass but, on the contrary, it increases substantially for initial masses up to $\rm 10~M_\odot$ and then flattens out for the more massive ones. This implies that the luminosity of the plateau at 30 days does not vary smoothly as well with the initial progenitor mass (see Figure \ref{fig:multilc}). Moreover, the initial mass-explosion energy relation adopted also explains the fact that the length of the plateau as a function of the progenitor mass, that should progressively increase, shows, on the contrary, an inversion moving from the $\rm 10~M_\odot$ to the $\rm 11~M_\odot$. The reason for such a behavior is that the substantially higher explosion energy of the $\rm 11~M_\odot$, compared to that of the $\rm 10~M_\odot$, makes the eulerian photosphere coordinate to recede faster in this model and therefore to approach the edge of the CO core, that marks the end of the plateau phase, earlier. It goes without saying that Figure \ref{fig:multilc} shows the light curves of these models computed with one possible choice of the various relevant parameters (e.g., explosion energy, $\rm ^{56}Ni$ ejected, extension and degree of mixing of the chemical composition, etc.). Different choices could be adopted in order to fit, case by case, the observed light curves of specific supernovae and transients in general.

\section{Summary and Conclusions}
We have presented and discussed the presupernova evolution of stars with initial mass 9.22, 10, 11, 12, 13, and $\rm 15~M_\odot$, from the pre-main sequence phase up to the onset of the iron core collapse, their hydrostatic and explosive nucleosynthesis and their associated bolometric supernova light curves. The main results of this study are the following:

(1) the density gradient in the region between the edge of the iron core and the He core at the presupernova stage steepens progressively as the initial mass decreases and turns almost vertical in the transition between the 10 and the $\rm 9.22~M_\odot$; this is due to the consequence of the second dredge up that occurs only in the lower mass model of this grid, i.e., the $\rm 9.22~M_\odot$ star;

(2) the yields of all the alpha elements (C, O, Ne, Mg, Si, S, Ar and Ca) decrease significantly as the initial mass decreases below $\rm \sim 15~M_\odot$ even by more than an order of magnitude;

(3) among the odd-Z elements, some of them (like, e.g., Na, Al and P) behave like the alpha elements while some others (like, e.g., N, F, K and Sc) show a much smoother decrease with decreasing the mass of the star;

(4) as a consequence the linear extrapolation of the ejected masses (in logarithm) of most of the isotopes down to the lower masses overestimates, even significantly, their total yields,
therefore the use of these yields is essential in galactic chemical evolution models;

(5) the different behavior found between all even-Z and some odd-Z elements can be used as a proxy for a supernova progenitor in the mass range presently studied; in particular a high value of the observed element abundance ratios [X/C,O,Mg] in the spectra of a core collapse supernova can be used to infer a low mass supernova progenitor;

(6) the elements heavier than Zn appear to decrease almost linearly (in logarithm) as the initial mass decreases;


(7) the production of the isotopes of the elements from O to Ca increases with increasing the mass of the star and roughly maintains a scaled solar distribution, with few exceptions;

(8) the ratio between the average overproduction of the isotopes from O to S and the one of the isotopes between Cl and C increases with the initial mass of the star;

(9) the production of the neutron capture elements from Ga to Zr increases with the mass of the stars such that a) the [weak-s/O] ratios scale inversely with the initial mass and b) for each one of these elements the more neutron rich isotopes are in general underproduced with respect to the less neutron rich ones; elements heavier than Zr are not produced at all, as expected;

(10) the contribution of stars in the range $\rm 9.22-13~M_\odot$ to the chemical enrichment of a generation of massive stars in the range $\rm 9.22-120~M_\odot$ is negligible for all the isotopes;

(12) the typical bolometric light curves expected from the explosion of stars in the range $\rm 9.22-15~M_\odot$ are characterized by a plateau phase with a luminosity at 30 days that ranges between ${\rm Log}~{(L/L_\odot)\sim 41.5 }$ and ${(L/L_\odot)\sim 42.1 }$ and with a duration in the range between $\sim 110$ and $\sim 140$ days.

Finally, we intend to examine the impact of stellar rotation on the results presented here. Rotation is known to have two primary effects: it leads to larger stellar cores due to rotation-induced mixing and increases the efficiency of mass loss \citep{cl13}. Consequently, we anticipate that our current scenario could change significantly when rotation is incorporated into the stellar models. A detailed quantitative analysis of how these effects influence the mass limits for forming ONeMg White Dwarfs, ECSNe, and CCSNe will be presented in a future publication.


\startlongtable
\begin{deluxetable*}{lcccccc}\label{tab_main_prop}
\tabletypesize{\small}
\tablecolumns{7}
\tablecaption{Main properties of the computed models}
\tablehead{\colhead{} & \colhead{9.22} & \colhead{10} & \colhead{11} & \colhead{12} & \colhead{13} & \colhead{15} } 
\startdata    
\cutinhead{ H burning}                                       
  $\rm M_{CC}$              & 3.33     & 3.71    & 4.19    & 4.74    & 5.24    & 6.39     \\   
  $\rm t$                   & 2.91(7)  & 2.50(7) & 2.11(7) & 1.83(7) & 1.62(7) & 1.31(7)  \\   
  $\rm M_{He }$             & 1.66     & 1.90    & 2.19    & 2.54    & 2.93    & 3.70     \\   
  $\rm M_{tot}$             & 9.18     & 9.96    & 10.91   & 11.89   & 12.88   & 14.89    \\   
\cutinhead{ He burning}                                                                 
  $\rm M_{CC}$              & 1.09     & 1.23    & 1.49    & 1.74    & 2.03    & 2.69     \\    
  $\rm t$                   & 2.50(6)  & 2.12(6) & 1.74(6) & 1.48(6) & 1.27(6) & 1.01(6)  \\   
  $\rm ^{12}C$              & 0.453    & 0.448   & 0.434   & 0.423   & 0.412   & 0.390    \\   
  $\rm M_{CO}$              & 1.08     & 1.23    & 1.48    & 1.74    & 2.03    & 2.69     \\   
  $\rm M_{He}$              & 2.51     & 2.72    & 3.10    & 3.49    & 3.90    & 4.77     \\   
  $\rm M_{CE}$              & 4.09     & 3.64    & 3.87    & 4.16    & 4.61    & 5.32     \\   
  $\rm M_{tot}$             & 8.89     & 9.61    & 10.39   & 11.28   & 12.15   & 13.93     \\   
\cutinhead{ Quantities at Carbon Ignition}                                                            
  $\rm Log(T_{ign})$        &    8.831  & 8.833 & 8.834 & 8.841 & 8.843 &    8.854  \\   
  $\rm Log(\rho_{ign})$     &    6.133  & 6.096 & 5.881 & 5.725 & 5.583 &    5.381  \\   
  $\rm \psi_{ign}$          &    2.450  & 2.272 & 1.412 & 0.832 & 0.374 &    -0.25  \\     
  $\rm M_{ign}$             &    0.082  & 0.000 & 0.000 & 0.000 & 0.000 &    0.000  \\   
  $\rm M_{CO}$              &    1.299  & 1.388 & 1.590 & 1.800 & 2.066 &    2.685  \\    
  $\rm M_{He}$              &    2.545  & 2.717 & 3.098 & 3.486 & 3.905 &    4.777  \\   
  $\rm M_{CE}$              &    2.548  & 2.817 & 3.218 & 3.627 & 4.065 &    4.985  \\   
  $\rm M_{tot}$             &    8.828  & 9.537 & 10.357& 11.212&12.017 &   13.684  \\
\cutinhead{ Quantities at Neon Ignition}                                                            
  $\rm ^{12}C    $              & 0.004   & 0.000   & 0.000   & 0.000   & 0.000   & 0.000   \\   
  $\rm M_{C}$                   & 0.000   & 0.000   & 0.000   & 0.000   & 0.000   & 0.000   \\   
  $\rm Log(T_{ign})$            & 9.138   & 9.179   & 9.160   & 9.166   & 9.167   & 9.145   \\   
  $\rm Log(\rho_{ign})$         & 7.443   & 7.324   & 7.279   & 7.227   & 7.197   & 6.990   \\   
  $\rm \psi_{ign}$              & 6.653   & 5.216   & 5.193   & 4.790   & 4.603   & 3.709   \\     
  $\rm M_{ign}$                 & 0.932   & 0.571   & 0.200   & 0.058   & 0.000   & 0.000   \\   
  $\rm M_{ONe}$                 & 1.35278 & 1.44708 & 1.58160 & 1.73310 & 1.96181 & 2.44923 \\
  $\rm M_{CO}$                  & 1.36596 & 1.47756 & 1.69028 & 1.87176 & 2.16383 & 2.76339 \\   
  $\rm M_{He}$                  & 1.57790 & 2.71881 & 3.10040 & 3.48811 & 3.90071 & 4.77717 \\   
  $\rm M_{CE}$                  & 1.57879 & 2.73556 & 3.16664 & 3.69772 & 4.02512 & 4.97346 \\   
  $\rm M_{tot}$                 & 8.754   & 9.42941 & 10.2382 & 11.0858 & 11.9091 & 13.6103 \\   
\cutinhead{ Quantities at Si-S Ignition}                                                            
 $\rm Log(T_{ign})$             & 9.509    &  9.473  & 9.475   & 9.494   & 9.530   & 9.513    \\   
 $\rm Log(\rho_{ign})$          & 8.194    &  8.900  & 8.448   & 9.032   & 8.305   & 8.501    \\   
 $\rm \psi_{ign}$               & 5.77     &  12.36  & 8.087   & 13.26   & 6.154   & 7.748    \\      
 $\rm M_{ign}$                  & 0.948    & 0.069   & 0.457   & 0.000   & 0.404   & 0.000    \\   
 $\rm M_{SiS}$                  & 1.32370  & 1.33520 & 1.41943 & 1.44115 & 1.48123 & 1.50483  \\   
 $\rm M_{ONe}$                  & 1.35702  & 1.46271 & 1.65755 & 1.84782 & 2.08227 & 2.63457  \\   
 $\rm M_{CO}$                   & 1.36596  & 1.47625 & 1.69221 & 1.90364 & 2.16669 & 2.76339  \\   
 $\rm M_{He}$                   & 1.57763  & 2.71881 & 3.10040 & 3.48811 & 3.90071 & 4.77717  \\   
 $\rm M_{CE}$                   & 1.57787  & 2.73420 & 3.16664 & 3.59772 & 4.02512 & 4.97346  \\   
 $\rm M_{tot}$                  & 8.754    & 9.429   & 10.2380 & 11.0857 & 11.9090  &13.6102   \\   
\cutinhead{ Quantities at Presupernova Stage}
 $\rm M_{Fe}$                   & 1.263    & 1.211   & 1.316   & 1.333   & 1.456   & 1.398    \\   
 $\rm Log(T_{c})$               & 9.684    & 9.720   & 9.755   & 9.804   & 9.755   & 9.772    \\   
 $\rm Log(\rho_{c})$            & 9.669    & 9.521   & 9.460   & 9.258   & 9.274   & 9.271    \\   
 $\rm \psi_{c}$                 & 14.79    & 11.59   & 10.07   & 7.458   & 8.532   & 8.177    \\      
 $\rm Y_{e,c}$                  & 0.437    & 0.438   & 0.436   & 0.440   & 0.438   & 0.438         
\enddata 
\end{deluxetable*}

\startlongtable
\floattable
\begin{deluxetable}{lcccccc}
\tablecolumns{7}
\tablecaption{Ejected Masses at $\rm 2.5\cdot 10^4~s$ \label{tabyields}}
\tablehead{$\rm Isotope$ & $\rm 9.22~M_\odot$ & $\rm 10~M_\odot$ & $\rm 11~M_\odot$ & $\rm 12~M_\odot$ & $\rm 13~M_\odot$ & $\rm 15~M_\odot$}
\startdata
$\rm^{1}H   $ &  4.6758E+00 & 5.0550E+00 & 5.4726E+00 & 5.8842E+00 & 6.2723E+00 & 7.0091E+00 \\    
$\rm^{2}H   $ &  1.2423E-06 & 1.3362E-06 & 2.7773E-06 & 3.1539E-06 & 3.6321E-06 & 3.4408E-06 \\    
$\rm^{3}H   $ &  7.5762E-22 & 4.8001E-16 & 3.7366E-16 & 3.5642E-33 & 1.0151E-30 & 6.7190E-27 \\    
$\rm^{3}He  $ &  3.0811E-04 & 3.2296E-04 & 3.3389E-04 & 3.4690E-04 & 3.6025E-04 & 3.9009E-04 \\    
$\rm^{4}He  $ &  3.0572E+00 & 3.3282E+00 & 3.6785E+00 & 4.0264E+00 & 4.3408E+00 & 4.9398E+00 \\    
$\rm^{6}Li  $ &  2.5150E-11 & 2.7051E-11 & 5.6228E-11 & 6.3853E-11 & 7.3535E-11 & 6.9662E-11 \\    
$\rm^{7}Li  $ &  3.5727E-10 & 3.8455E-10 & 8.0077E-10 & 9.0700E-10 & 1.0445E-09 & 9.8952E-10 \\    
$\rm^{7}Be  $ &  2.5805E-14 & 2.3128E-11 & 4.6408E-10 & 4.8469E-22 & 2.1430E-20 & 1.4975E-17 \\    
$\rm^{9}Be  $ &  1.4844E-11 & 1.7597E-11 & 1.9878E-11 & 2.1469E-11 & 2.3268E-11 & 2.4634E-11 \\    
$\rm^{10}Be $ &  1.7661E-18 & 2.9197E-19 & 1.9306E-19 & 9.6215E-40 & 1.0406E-39 & 1.2051E-39 \\    
$\rm^{10}B  $ &  1.7214E-10 & 2.0570E-10 & 2.2212E-10 & 2.3666E-10 & 2.5033E-10 & 2.7643E-10 \\    
$\rm^{11}B  $ &  5.4881E-09 & 6.0971E-09 & 6.6036E-09 & 7.0917E-09 & 7.5589E-09 & 8.4702E-09 \\    
$\rm^{12}C  $ &  2.7650E-02 & 3.6902E-02 & 5.4731E-02 & 7.9709E-02 & 1.1326E-01 & 1.7461E-01 \\    
$\rm^{13}C  $ &  4.3668E-04 & 4.9003E-04 & 5.3509E-04 & 5.8725E-04 & 6.4081E-04 & 7.2933E-04 \\    
$\rm^{14}C  $ &  2.7811E-08 & 5.4103E-08 & 1.5229E-08 & 2.2410E-08 & 8.8157E-08 & 2.0197E-07 \\    
$\rm^{13}N  $ &  6.7753E-15 & 2.3873E-15 & 1.8533E-14 & 5.3464E-14 & 5.7210E-14 & 3.6098E-14 \\    
$\rm^{14}N  $ &  2.4452E-02 & 2.4141E-02 & 2.9164E-02 & 3.2127E-02 & 3.4300E-02 & 3.7577E-02 \\    
$\rm^{15}N  $ &  5.2783E-06 & 6.2059E-06 & 5.8130E-06 & 6.2593E-06 & 6.6118E-06 & 7.3764E-06 \\    
$\rm^{16}N  $ &  3.8018E-30 & 1.4585E-31 & 6.2176E-28 & 4.2320E-30 & 5.1761E-27 & 2.9792E-33 \\    
$\rm^{15}O  $ &  1.3235E-21 & 2.7337E-17 & 2.7620E-17 & 3.7233E-17 & 1.5500E-15 & 5.3882E-16 \\    
$\rm^{16}O  $ &  4.8055E-02 & 7.4294E-02 & 1.5758E-01 & 2.2866E-01 & 3.5156E-01 & 5.9481E-01 \\    
$\rm^{17}O  $ &  5.4097E-05 & 5.4178E-05 & 5.5634E-05 & 5.7930E-05 & 5.7043E-05 & 5.7326E-05 \\    
$\rm^{18}O  $ &  1.9470E-04 & 2.7384E-03 & 9.8078E-04 & 7.0057E-04 & 9.3260E-04 & 1.4677E-03 \\    
$\rm^{19}O  $ &  1.2301E-25 & 3.2049E-25 & 4.8739E-21 & 1.1491E-23 & 3.1449E-21 & 1.8962E-27 \\    
$\rm^{17}F  $ &  3.4269E-22 & 1.8867E-24 & 6.3065E-23 & 8.8326E-27 & 1.4790E-24 & 4.6469E-27 \\    
$\rm^{18}F  $ &  7.1406E-11 & 4.2976E-08 & 5.7844E-09 & 3.2255E-09 & 8.4228E-09 & 4.1860E-09 \\    
$\rm^{19}F  $ &  3.2150E-06 & 3.1711E-06 & 3.2536E-06 & 3.4731E-06 & 3.7473E-06 & 4.1927E-06 \\    
$\rm^{20}F  $ &  7.9977E-25 & 5.6514E-25 & 2.0108E-24 & 2.3746E-25 & 1.0332E-21 & 7.3223E-26 \\    
$\rm^{20}Ne $ &  1.1675E-02 & 1.9009E-02 & 3.5053E-02 & 7.8528E-02 & 8.9161E-02 & 2.9002E-01 \\    
$\rm^{21}Ne $ &  4.2647E-05 & 5.0157E-05 & 5.4575E-05 & 7.5797E-05 & 9.4960E-05 & 2.1879E-04 \\    
$\rm^{22}Ne $ &  2.6311E-03 & 3.6438E-03 & 2.9541E-03 & 3.8325E-03 & 5.0114E-03 & 9.1615E-03 \\    
$\rm^{23}Ne $ &  3.6684E-18 & 1.5261E-17 & 5.2785E-17 & 7.7697E-18 & 4.3672E-15 & 1.0989E-19 \\    
$\rm^{21}Na $ &  9.0210E-18 & 3.7789E-17 & 1.3938E-16 & 1.8930E-17 & 2.2682E-16 & 2.4240E-21 \\    
$\rm^{22}Na $ &  1.8750E-08 & 2.8603E-08 & 3.8205E-08 & 2.2876E-07 & 2.3541E-07 & 1.0221E-06 \\    
$\rm^{23}Na $ &  5.6381E-04 & 6.4895E-04 & 6.7919E-04 & 8.6961E-04 & 1.2948E-03 & 3.7936E-03 \\    
$\rm^{24}Na $ &  3.6171E-08 & 3.1020E-08 & 2.6988E-09 & 1.8456E-08 & 4.5289E-08 & 3.6129E-07 \\    
$\rm^{23}Mg $ &  1.7170E-17 & 9.1369E-17 & 2.4084E-16 & 7.6404E-17 & 7.6902E-14 & 5.3389E-14 \\    
$\rm^{24}Mg $ &  5.0809E-03 & 7.8729E-03 & 2.7018E-02 & 3.8297E-02 & 5.1547E-02 & 8.8128E-02 \\    
$\rm^{25}Mg $ &  5.8413E-04 & 8.0429E-04 & 1.6962E-03 & 2.9684E-03 & 3.3399E-03 & 8.2973E-03 \\    
$\rm^{26}Mg $ &  9.0947E-04 & 1.1796E-03 & 1.8271E-03 & 3.1698E-03 & 3.4287E-03 & 8.3621E-03 \\    
$\rm^{27}Mg $ &  6.4071E-15 & 5.0506E-14 & 1.5113E-14 & 6.3495E-14 & 2.2675E-11 & 1.5641E-14 \\    
$\rm^{25}Al $ &  1.5195E-18 & 5.7726E-20 & 5.6904E-19 & 4.3594E-19 & 4.9295E-18 & 1.0693E-21 \\    
$\rm^{26}Al $ &  2.6652E-06 & 5.2074E-06 & 1.5387E-05 & 3.1272E-05 & 2.5040E-05 & 8.3182E-05 \\    
$\rm^{27}Al $ &  5.3350E-04 & 8.4107E-04 & 2.8805E-03 & 4.0425E-03 & 5.8592E-03 & 9.3676E-03 \\    
$\rm^{28}Al $ &  3.3686E-16 & 1.8411E-14 & 1.2209E-14 & 1.6189E-13 & 1.0429E-10 & 2.4681E-11 \\    
$\rm^{27}Si $ &  1.9359E-22 & 7.0198E-20 & 3.4703E-18 & 1.6946E-18 & 2.4832E-14 & 1.9359E-14 \\    
$\rm^{28}Si $ &  1.1218E-02 & 2.0415E-02 & 4.2969E-02 & 5.6178E-02 & 9.5674E-02 & 1.0953E-01 \\    
$\rm^{29}Si $ &  3.0333E-04 & 4.7386E-04 & 1.9837E-03 & 2.3388E-03 & 3.7150E-03 & 4.1090E-03 \\    
$\rm^{30}Si $ &  2.5696E-04 & 5.2411E-04 & 2.6674E-03 & 2.7305E-03 & 5.4616E-03 & 4.8362E-03 \\    
$\rm^{31}Si $ &  7.4173E-10 & 2.9594E-09 & 1.6268E-08 & 1.9627E-08 & 3.8702E-08 & 1.2930E-07 \\    
$\rm^{32}Si $ &  1.9599E-10 & 4.1728E-10 & 1.7022E-08 & 9.3552E-09 & 2.3476E-08 & 2.3220E-08 \\    
$\rm^{29}P  $ &  3.2696E-21 & 4.8094E-22 & 7.1248E-22 & 7.9198E-23 & 2.7038E-20 & 6.4583E-23 \\    
$\rm^{30}P  $ &  1.0818E-16 & 1.4838E-13 & 1.8084E-13 & 2.6224E-13 & 4.9615E-11 & 6.2495E-12 \\    
$\rm^{31}P  $ &  6.6814E-05 & 1.4582E-04 & 5.2836E-04 & 4.8062E-04 & 1.3169E-03 & 9.1697E-04 \\    
$\rm^{32}P  $ &  2.4533E-08 & 9.1148E-08 & 1.1072E-06 & 1.2300E-06 & 3.3510E-06 & 2.7591E-06 \\    
$\rm^{33}P  $ &  1.6430E-08 & 6.0784E-08 & 4.0613E-07 & 4.1477E-07 & 1.0071E-06 & 8.4029E-07 \\    
$\rm^{34}P  $ &  4.8144E-24 & 1.8971E-24 & 6.0519E-23 & 4.7087E-23 & 7.7311E-19 & 4.2050E-19 \\    
$\rm^{31}S  $ &  9.2194E-29 & 5.8417E-24 & 4.2618E-22 & 1.5808E-22 & 2.6854E-17 & 2.3891E-17 \\    
$\rm^{32}S  $ &  5.6091E-03 & 9.7466E-03 & 1.4751E-02 & 2.0675E-02 & 3.2052E-02 & 3.9790E-02 \\    
$\rm^{33}S  $ &  3.3116E-05 & 7.8525E-05 & 1.2429E-04 & 1.7143E-04 & 3.0517E-04 & 3.1812E-04 \\    
$\rm^{34}S  $ &  2.0560E-04 & 5.7826E-04 & 8.8788E-04 & 1.2538E-03 & 2.2569E-03 & 2.1916E-03 \\    
$\rm^{35}S  $ &  2.8062E-08 & 1.1968E-07 & 1.0336E-06 & 4.9299E-07 & 4.3798E-06 & 1.3777E-06 \\    
$\rm^{36}S  $ &  5.7513E-07 & 6.6469E-07 & 1.9077E-06 & 1.2682E-06 & 5.0769E-06 & 2.8162E-06 \\    
$\rm^{37}S  $ &  1.5642E-20 & 1.9448E-19 & 3.0568E-20 & 9.4104E-21 & 8.6710E-17 & 1.1529E-20 \\    
$\rm^{33}Cl $ &  2.9675E-22 & 1.3806E-24 & 3.3096E-24 & 9.6377E-25 & 1.7453E-23 & 1.5195E-26 \\    
$\rm^{34}Cl $ &  1.8579E-23 & 2.4123E-26 & 6.2131E-26 & 1.0951E-26 & 1.5450E-21 & 2.5785E-21 \\    
$\rm^{35}Cl $ &  5.4683E-05 & 8.2636E-05 & 1.1296E-04 & 1.2363E-04 & 2.6842E-04 & 2.0332E-04 \\    
$\rm^{36}Cl $ &  9.4946E-08 & 2.6718E-07 & 9.1412E-07 & 7.6890E-07 & 3.3896E-06 & 1.7395E-06 \\    
$\rm^{37}Cl $ &  1.7510E-05 & 1.9679E-05 & 2.3209E-05 & 2.7193E-05 & 3.2719E-05 & 4.5875E-05 \\    
$\rm^{38}Cl $ &  2.7666E-13 & 4.3818E-13 & 4.2430E-13 & 1.0857E-12 & 5.7297E-12 & 1.4707E-11 \\    
$\rm^{36}Ar $ &  1.1084E-03 & 1.8276E-03 & 2.5955E-03 & 3.6123E-03 & 5.0916E-03 & 6.6016E-03 \\    
$\rm^{37}Ar $ &  9.1076E-07 & 3.1905E-06 & 3.7848E-06 & 5.7340E-06 & 9.2070E-06 & 1.1486E-05 \\    
$\rm^{38}Ar $ &  1.2272E-04 & 2.7007E-04 & 3.3911E-04 & 4.5031E-04 & 7.0338E-04 & 8.0489E-04 \\    
$\rm^{39}Ar $ &  6.5042E-08 & 1.3063E-07 & 3.2235E-07 & 5.5508E-07 & 7.7296E-07 & 1.4666E-06 \\    
$\rm^{40}Ar $ &  1.9152E-07 & 2.3218E-07 & 3.2658E-07 & 4.4016E-07 & 5.2363E-07 & 7.2122E-07 \\    
$\rm^{41}Ar $ &  1.1506E-11 & 7.4178E-12 & 5.5285E-12 & 1.7661E-11 & 4.4620E-11 & 2.0955E-10 \\    
$\rm^{37}K  $ &  8.0646E-28 & 5.0971E-33 & 3.7840E-33 & 1.1576E-33 & 1.1526E-32 & 4.4047E-34 \\    
$\rm^{38}K  $ &  1.3264E-17 & 9.6610E-15 & 2.4265E-15 & 2.5977E-15 & 2.3378E-14 & 1.6733E-14 \\    
$\rm^{39}K  $ &  2.5621E-05 & 3.5635E-05 & 4.1645E-05 & 5.0494E-05 & 6.3429E-05 & 7.8808E-05 \\    
$\rm^{40}K  $ &  7.3989E-08 & 1.0056E-07 & 1.6388E-07 & 2.2466E-07 & 3.7666E-07 & 5.9375E-07 \\    
$\rm^{41}K  $ &  1.7407E-06 & 1.9017E-06 & 2.1364E-06 & 2.4156E-06 & 2.6525E-06 & 3.4532E-06 \\    
$\rm^{42}K  $ &  4.7916E-10 & 5.1587E-10 & 4.8662E-10 & 3.8446E-09 & 2.0201E-09 & 3.4300E-08 \\    
$\rm^{40}Ca $ &  1.1158E-03 & 1.7535E-03 & 2.4400E-03 & 3.2788E-03 & 4.3963E-03 & 5.7382E-03 \\    
$\rm^{41}Ca $ &  5.5133E-07 & 1.0964E-06 & 1.5427E-06 & 2.1665E-06 & 2.8652E-06 & 4.8191E-06 \\    
$\rm^{42}Ca $ &  4.7843E-06 & 9.1925E-06 & 1.1501E-05 & 1.4950E-05 & 2.0373E-05 & 2.8859E-05 \\    
$\rm^{43}Ca $ &  8.1914E-07 & 9.9013E-07 & 1.3543E-06 & 1.7381E-06 & 2.1332E-06 & 3.7897E-06 \\    
$\rm^{44}Ca $ &  1.1914E-05 & 1.3105E-05 & 1.4975E-05 & 1.6933E-05 & 1.8956E-05 & 2.4321E-05 \\    
$\rm^{45}Ca $ &  8.7525E-09 & 9.0277E-09 & 6.1351E-08 & 1.4886E-07 & 3.7436E-07 & 4.8672E-07 \\    
$\rm^{46}Ca $ &  2.9356E-08 & 4.2805E-08 & 7.8043E-08 & 1.0319E-07 & 2.4824E-07 & 1.4624E-07 \\    
$\rm^{47}Ca $ &  9.7176E-11 & 5.2856E-11 & 1.1974E-10 & 7.5045E-10 & 1.1219E-09 & 2.9825E-09 \\    
$\rm^{48}Ca $ &  1.1369E-06 & 1.2350E-06 & 1.3566E-06 & 1.4826E-06 & 1.5949E-06 & 1.8557E-06 \\    
$\rm^{49}Ca $ &  2.3685E-19 & 1.8911E-18 & 4.2564E-19 & 2.8026E-19 & 4.4358E-17 & 6.5798E-19 \\    
$\rm^{41}Sc $ &  1.5783E-35 & 1.4906E-44 & 1.3813E-44 & 1.5372E-44 & 1.8884E-44 & 2.2468E-44 \\    
$\rm^{42}Sc $ &  1.1283E-30 & 2.4893E-38 & 7.1411E-40 & 5.6740E-42 & 7.3924E-37 & 2.3762E-36 \\    
$\rm^{43}Sc $ &  4.4859E-09 & 1.4417E-08 & 2.9983E-08 & 2.6967E-08 & 1.4422E-08 & 1.8290E-08 \\    
$\rm^{44}Sc $ &  1.8839E-10 & 7.2320E-10 & 1.1852E-09 & 1.7830E-09 & 2.3165E-09 & 5.0630E-09 \\    
$\rm^{45}Sc $ &  4.3952E-07 & 5.3659E-07 & 7.7653E-07 & 9.1558E-07 & 1.2117E-06 & 1.7701E-06 \\    
$\rm^{46}Sc $ &  5.2486E-09 & 7.0962E-09 & 2.5481E-08 & 7.0997E-08 & 1.2811E-07 & 2.3156E-07 \\    
$\rm^{47}Sc $ &  1.9446E-09 & 1.8635E-09 & 3.9259E-09 & 1.4664E-08 & 1.3665E-08 & 5.8944E-08 \\    
$\rm^{48}Sc $ &  4.3756E-10 & 4.1966E-10 & 7.8319E-10 & 3.4768E-09 & 1.5202E-09 & 1.7128E-08 \\    
$\rm^{49}Sc $ &  3.4832E-13 & 4.9919E-13 & 1.3754E-12 & 3.0317E-12 & 2.1351E-12 & 3.0784E-11 \\    
$\rm^{44}Ti $ &  9.3365E-07 & 2.3101E-06 & 3.2364E-06 & 3.1326E-06 & 2.9069E-06 & 3.8157E-06 \\    
$\rm^{45}Ti $ &  3.7187E-09 & 7.1715E-09 & 9.8911E-09 & 1.2729E-08 & 1.4179E-08 & 2.2051E-08 \\    
$\rm^{46}Ti $ &  2.6699E-06 & 4.3276E-06 & 5.7666E-06 & 6.9586E-06 & 8.8453E-06 & 1.2212E-05 \\    
$\rm^{47}Ti $ &  1.8832E-06 & 2.1323E-06 & 2.5285E-06 & 2.8025E-06 & 3.2095E-06 & 3.9369E-06 \\    
$\rm^{48}Ti $ &  1.8270E-05 & 1.9874E-05 & 2.2003E-05 & 2.3986E-05 & 2.5996E-05 & 2.9811E-05 \\    
$\rm^{49}Ti $ &  1.4762E-06 & 1.6255E-06 & 1.9038E-06 & 2.1917E-06 & 2.5014E-06 & 3.2227E-06 \\    
$\rm^{50}Ti $ &  1.4538E-06 & 1.6339E-06 & 2.0452E-06 & 2.4629E-06 & 2.9448E-06 & 4.2319E-06 \\    
$\rm^{51}Ti $ &  4.9874E-17 & 1.4795E-16 & 7.1924E-17 & 3.7077E-16 & 5.9595E-14 & 1.6622E-17 \\    
$\rm^{45}V  $ &  1.0173E-43 & 1.4389E-44 & 1.7328E-44 & 1.9654E-44 & 1.4465E-43 & 5.5019E-44 \\    
$\rm^{46}V  $ &  4.1943E-45 & 8.3539E-45 & 1.0071E-44 & 1.1412E-44 & 6.7669E-42 & 1.1329E-41 \\    
$\rm^{47}V  $ &  8.2455E-12 & 1.8111E-11 & 3.1726E-11 & 2.9901E-11 & 2.7389E-11 & 3.8005E-11 \\    
$\rm^{48}V  $ &  1.4429E-08 & 2.9110E-08 & 4.1858E-08 & 5.1949E-08 & 7.5042E-08 & 1.0192E-07 \\    
$\rm^{49}V  $ &  1.4869E-06 & 1.7224E-06 & 2.6726E-06 & 3.5648E-06 & 5.5605E-06 & 7.3232E-06 \\    
$\rm^{50}V  $ &  9.7472E-09 & 1.4859E-08 & 2.7714E-08 & 3.7753E-08 & 6.7469E-08 & 8.9528E-08 \\    
$\rm^{51}V  $ &  2.5381E-06 & 2.7669E-06 & 3.1040E-06 & 3.4029E-06 & 3.7529E-06 & 4.3585E-06 \\    
$\rm^{52}V  $ &  1.8645E-17 & 8.0168E-17 & 6.7932E-17 & 2.1478E-16 & 4.5943E-14 & 6.0760E-15 \\    
$\rm^{48}Cr $ &  1.2072E-05 & 2.2732E-05 & 3.2891E-05 & 3.7259E-05 & 5.2867E-05 & 6.4146E-05 \\    
$\rm^{49}Cr $ &  2.8645E-09 & 3.5792E-09 & 5.4527E-09 & 7.1979E-09 & 1.1333E-08 & 1.5008E-08 \\    
$\rm^{50}Cr $ &  2.3449E-05 & 1.9837E-05 & 3.4453E-05 & 4.4968E-05 & 7.6946E-05 & 1.2027E-04 \\    
$\rm^{51}Cr $ &  4.5198E-06 & 4.1578E-06 & 7.0512E-06 & 9.7100E-06 & 1.5713E-05 & 2.1080E-05 \\    
$\rm^{52}Cr $ &  1.1103E-04 & 1.2217E-04 & 1.3979E-04 & 1.5190E-04 & 1.7280E-04 & 1.9074E-04 \\    
$\rm^{53}Cr $ &  1.2638E-05 & 1.3730E-05 & 1.5181E-05 & 1.6612E-05 & 1.8011E-05 & 2.0966E-05 \\    
$\rm^{54}Cr $ &  3.5006E-06 & 3.9630E-06 & 5.0048E-06 & 6.0655E-06 & 7.4179E-06 & 1.0962E-05 \\    
$\rm^{55}Cr $ &  3.7380E-17 & 9.3991E-17 & 7.3668E-17 & 2.2325E-16 & 1.0622E-13 & 1.8200E-18 \\    
$\rm^{50}Mn $ &  2.7896E-45 & 5.6328E-45 & 6.7850E-45 & 7.6971E-45 & 9.2335E-45 & 1.1211E-44 \\    
$\rm^{51}Mn $ &  1.2106E-08 & 1.1651E-08 & 1.9430E-08 & 2.6791E-08 & 4.3548E-08 & 5.9754E-08 \\    
$\rm^{52}Mn $ &  4.5255E-05 & 7.4887E-05 & 1.1350E-04 & 1.3898E-04 & 2.2633E-04 & 2.8438E-04 \\    
$\rm^{53}Mn $ &  3.1311E-05 & 3.6165E-05 & 5.7841E-05 & 7.7756E-05 & 1.2600E-04 & 1.6286E-04 \\    
$\rm^{54}Mn $ &  7.2374E-09 & 2.1554E-08 & 6.3528E-08 & 8.1462E-08 & 1.4833E-07 & 1.6143E-07 \\    
$\rm^{55}Mn $ &  8.7178E-05 & 9.4752E-05 & 1.0460E-04 & 1.1421E-04 & 1.2346E-04 & 1.4263E-04 \\    
$\rm^{56}Mn $ &  3.9803E-10 & 1.8351E-09 & 2.5389E-09 & 7.2585E-09 & 7.5354E-09 & 1.9713E-08 \\    
$\rm^{57}Mn $ &  6.8692E-23 & 4.1748E-21 & 4.2709E-20 & 2.8315E-18 & 3.1343E-16 & 5.5460E-18 \\    
$\rm^{52}Fe $ &  1.2664E-04 & 2.1010E-04 & 3.1832E-04 & 3.8954E-04 & 6.3399E-04 & 7.9628E-04 \\    
$\rm^{53}Fe $ &  2.2704E-17 & 2.0197E-14 & 2.6534E-14 & 2.5817E-14 & 1.7087E-13 & 1.0834E-13 \\    
$\rm^{54}Fe $ &  2.0026E-03 & 1.7955E-03 & 2.9912E-03 & 4.0753E-03 & 6.6620E-03 & 9.7541E-03 \\    
$\rm^{55}Fe $ &  3.3100E-05 & 3.4437E-05 & 5.7615E-05 & 8.0443E-05 & 1.3247E-04 & 1.6833E-04 \\    
$\rm^{56}Fe $ &  9.3929E-03 & 1.0205E-02 & 1.1223E-02 & 1.2207E-02 & 1.3185E-02 & 1.5039E-02 \\    
$\rm^{57}Fe $ &  2.2685E-04 & 2.4896E-04 & 2.8886E-04 & 3.2951E-04 & 3.7298E-04 & 4.8761E-04 \\    
$\rm^{58}Fe $ &  3.7352E-05 & 5.1147E-05 & 9.9319E-05 & 1.4076E-04 & 2.1403E-04 & 3.7795E-04 \\    
$\rm^{59}Fe $ &  1.9694E-07 & 3.1119E-07 & 5.4301E-07 & 6.8852E-06 & 6.7279E-06 & 1.4568E-05 \\    
$\rm^{60}Fe $ &  1.2094E-07 & 3.7961E-07 & 1.3572E-06 & 2.2052E-06 & 6.2122E-06 & 3.3987E-06 \\    
$\rm^{61}Fe $ &  1.7599E-17 & 1.5776E-16 & 7.5776E-17 & 2.3414E-17 & 1.2781E-14 & 2.5756E-19 \\    
$\rm^{54}Co $ &  1.9121E-45 & 3.8609E-45 & 4.6508E-45 & 5.2761E-45 & 6.3292E-45 & 7.6846E-45 \\    
$\rm^{55}Co $ &  1.4973E-04 & 1.4403E-04 & 2.2760E-04 & 3.2710E-04 & 5.1980E-04 & 6.9525E-04 \\    
$\rm^{56}Co $ &  9.1307E-07 & 7.1665E-07 & 1.2873E-06 & 1.6633E-06 & 2.5501E-06 & 3.9613E-06 \\    
$\rm^{57}Co $ &  2.7431E-05 & 3.9748E-05 & 6.4177E-05 & 7.8189E-05 & 8.8377E-05 & 1.0060E-04 \\    
$\rm^{58}Co $ &  1.9682E-08 & 1.6935E-07 & 6.5496E-07 & 1.0385E-06 & 1.6274E-06 & 2.0502E-06 \\    
$\rm^{59}Co $ &  3.6130E-05 & 4.4922E-05 & 7.2536E-05 & 9.1920E-05 & 1.2371E-04 & 2.1757E-04 \\    
$\rm^{60}Co $ &  1.4363E-07 & 4.1280E-07 & 9.1859E-07 & 4.4312E-06 & 4.9808E-06 & 1.0954E-05 \\    
$\rm^{61}Co $ &  6.1228E-09 & 1.4088E-08 & 1.8885E-08 & 4.7726E-08 & 4.9217E-08 & 3.6728E-07 \\    
$\rm^{56}Ni $ &  1.0400E-02 & 1.9500E-02 & 2.9200E-02 & 3.5500E-02 & 4.1700E-02 & 5.4200E-02 \\    
$\rm^{57}Ni $ &  4.7359E-04 & 6.7861E-04 & 1.0800E-03 & 1.3064E-03 & 1.4423E-03 & 1.6420E-03 \\    
$\rm^{58}Ni $ &  1.7379E-03 & 1.6219E-03 & 3.1684E-03 & 4.0368E-03 & 5.0289E-03 & 4.8085E-03 \\    
$\rm^{59}Ni $ &  2.6333E-05 & 3.5148E-05 & 6.0601E-05 & 6.5564E-05 & 6.5900E-05 & 6.8971E-05 \\    
$\rm^{60}Ni $ &  2.7381E-04 & 4.8911E-04 & 7.1084E-04 & 7.2051E-04 & 6.9238E-04 & 8.8756E-04 \\    
$\rm^{61}Ni $ &  1.3192E-05 & 2.0500E-05 & 3.3992E-05 & 4.0509E-05 & 4.8072E-05 & 7.1574E-05 \\    
$\rm^{62}Ni $ &  3.0304E-05 & 4.1885E-05 & 6.7042E-05 & 8.7651E-05 & 1.1869E-04 & 1.7071E-04 \\    
$\rm^{63}Ni $ &  8.2361E-07 & 2.3386E-06 & 7.3613E-06 & 1.2519E-05 & 1.7901E-05 & 2.5519E-05 \\    
$\rm^{64}Ni $ &  9.2769E-06 & 1.1763E-05 & 1.9007E-05 & 2.6757E-05 & 3.6471E-05 & 4.4718E-05 \\    
$\rm^{65}Ni $ &  4.5876E-09 & 1.5463E-09 & 1.1666E-09 & 3.6727E-09 & 1.6962E-08 & 2.2069E-08 \\    
$\rm^{57}Cu $ &  2.3962E-45 & 4.8384E-45 & 5.8282E-45 & 6.6118E-45 & 7.9315E-45 & 9.6300E-45 \\    
$\rm^{58}Cu $ &  3.1178E-44 & 1.0109E-36 & 1.9802E-35 & 9.9990E-36 & 1.0053E-32 & 9.4990E-33 \\    
$\rm^{59}Cu $ &  4.3036E-25 & 8.7369E-22 & 1.3007E-22 & 1.9323E-22 & 1.4755E-19 & 3.6694E-20 \\    
$\rm^{60}Cu $ &  2.7000E-09 & 8.5065E-09 & 1.3226E-08 & 1.2619E-08 & 1.1115E-08 & 1.4199E-08 \\    
$\rm^{61}Cu $ &  3.5372E-06 & 7.1709E-06 & 1.2988E-05 & 1.2379E-05 & 1.1317E-05 & 1.1900E-05 \\    
$\rm^{62}Cu $ &  1.3274E-07 & 1.8444E-07 & 4.4252E-07 & 4.4318E-07 & 4.6790E-07 & 4.1846E-07 \\    
$\rm^{63}Cu $ &  5.3999E-06 & 6.1556E-06 & 6.8178E-06 & 8.1882E-06 & 9.5350E-06 & 1.5692E-05 \\    
$\rm^{64}Cu $ &  8.5162E-08 & 1.3204E-07 & 1.4733E-07 & 3.2531E-07 & 3.4235E-07 & 9.6541E-07 \\    
$\rm^{65}Cu $ &  3.2830E-06 & 4.0308E-06 & 6.3698E-06 & 8.9955E-06 & 1.1703E-05 & 1.9608E-05 \\    
$\rm^{66}Cu $ &  3.6567E-15 & 9.5292E-15 & 4.4760E-15 & 1.8364E-14 & 2.7360E-12 & 1.0291E-15 \\    
$\rm^{60}Zn $ &  2.1383E-25 & 1.6003E-23 & 9.1609E-24 & 1.4344E-23 & 2.0833E-23 & 3.8505E-23 \\    
$\rm^{61}Zn $ &  7.9428E-27 & 2.5343E-23 & 4.3982E-25 & 1.3081E-24 & 1.6793E-22 & 9.2779E-23 \\    
$\rm^{62}Zn $ &  9.6628E-05 & 1.3908E-04 & 3.3253E-04 & 3.3201E-04 & 3.4985E-04 & 3.0809E-04 \\    
$\rm^{63}Zn $ &  1.9447E-10 & 2.9091E-10 & 6.3118E-10 & 5.6853E-10 & 5.3945E-10 & 4.4006E-10 \\    
$\rm^{64}Zn $ &  7.4075E-06 & 9.2261E-06 & 1.0770E-05 & 1.1923E-05 & 1.2554E-05 & 1.8200E-05 \\    
$\rm^{65}Zn $ &  1.5897E-07 & 3.3925E-07 & 5.0893E-07 & 7.1979E-07 & 7.7855E-07 & 1.9667E-06 \\    
$\rm^{66}Zn $ &  5.6706E-06 & 7.7168E-06 & 9.3286E-06 & 1.2813E-05 & 1.5560E-05 & 2.4844E-05 \\    
$\rm^{67}Zn $ &  8.4195E-07 & 9.3483E-07 & 1.5701E-06 & 2.0935E-06 & 2.6353E-06 & 4.5403E-06 \\    
$\rm^{68}Zn $ &  3.8882E-06 & 4.5184E-06 & 7.4412E-06 & 9.4092E-06 & 1.3057E-05 & 1.7874E-05 \\    
$\rm^{69}Zn $ &  1.9392E-10 & 1.0464E-10 & 1.4288E-10 & 3.4045E-10 & 7.7626E-10 & 3.1014E-09 \\    
$\rm^{70}Zn $ &  1.4151E-07 & 1.1705E-07 & 1.7110E-07 & 2.4879E-07 & 3.1949E-07 & 4.8628E-07 \\    
$\rm^{71}Zn $ &  4.6210E-22 & 6.9235E-20 & 1.0466E-20 & 1.4585E-18 & 3.1325E-15 & 2.9116E-21 \\    
$\rm^{62}Ga $ &  1.1601E-45 & 2.3424E-45 & 2.8217E-45 & 3.2011E-45 & 3.8400E-45 & 4.6624E-45 \\    
$\rm^{63}Ga $ &  1.3322E-37 & 4.5119E-32 & 1.4560E-33 & 1.6324E-32 & 4.8808E-29 & 1.3010E-30 \\    
$\rm^{64}Ga $ &  2.2176E-27 & 8.4707E-25 & 1.3746E-28 & 2.6225E-27 & 2.9535E-25 & 2.1720E-25 \\    
$\rm^{65}Ga $ &  9.8671E-15 & 3.6323E-14 & 6.2073E-14 & 5.3081E-14 & 4.3904E-14 & 4.7807E-14 \\    
$\rm^{66}Ga $ &  3.6395E-07 & 7.0500E-07 & 1.7695E-06 & 1.6098E-06 & 1.5165E-06 & 1.3455E-06 \\    
$\rm^{67}Ga $ &  1.1893E-08 & 1.2633E-08 & 1.7173E-08 & 2.2146E-08 & 2.1531E-08 & 3.5424E-08 \\    
$\rm^{68}Ga $ &  1.0779E-10 & 1.0778E-10 & 9.9906E-11 & 1.7631E-10 & 1.8345E-10 & 4.3049E-10 \\    
$\rm^{69}Ga $ &  4.8993E-07 & 6.0946E-07 & 9.3487E-07 & 1.2199E-06 & 1.4532E-06 & 2.3587E-06 \\    
$\rm^{70}Ga $ &  1.0337E-13 & 5.2962E-13 & 9.6163E-13 & 1.2581E-12 & 1.0963E-11 & 3.0219E-12 \\    
$\rm^{71}Ga $ &  3.3334E-07 & 3.5461E-07 & 4.8371E-07 & 6.6704E-07 & 8.0185E-07 & 1.3630E-06 \\    
$\rm^{72}Ga $ &  8.3083E-09 & 2.2210E-09 & 1.2078E-09 & 4.2881E-09 & 7.8376E-09 & 1.9180E-08 \\    
$\rm^{64}Ge $ &  1.8577E-37 & 1.2738E-33 & 1.0082E-33 & 2.9586E-35 & 1.0103E-33 & 5.2879E-34 \\    
$\rm^{65}Ge $ &  1.4370E-43 & 1.1397E-35 & 8.5749E-37 & 2.1548E-36 & 2.0853E-33 & 3.6905E-33 \\    
$\rm^{66}Ge $ &  7.0592E-07 & 1.3741E-06 & 3.4406E-06 & 3.1290E-06 & 2.9429E-06 & 2.6088E-06 \\    
$\rm^{67}Ge $ &  3.5407E-15 & 8.0740E-15 & 1.9361E-14 & 1.5823E-14 & 1.3167E-14 & 1.0791E-14 \\    
$\rm^{68}Ge $ &  3.3982E-09 & 3.7913E-09 & 6.0787E-09 & 7.2849E-09 & 7.4554E-09 & 1.0397E-08 \\    
$\rm^{69}Ge $ &  5.9254E-10 & 8.3220E-10 & 4.2508E-10 & 1.0319E-09 & 1.0016E-09 & 2.3660E-09 \\    
$\rm^{70}Ge $ &  6.6209E-07 & 9.4553E-07 & 1.1466E-06 & 1.4963E-06 & 1.7202E-06 & 2.7569E-06 \\    
$\rm^{71}Ge $ &  2.8373E-08 & 2.8211E-08 & 9.9821E-08 & 1.6379E-07 & 2.0882E-07 & 3.6702E-07 \\    
$\rm^{72}Ge $ &  7.8501E-07 & 1.0185E-06 & 1.4602E-06 & 1.8232E-06 & 2.1951E-06 & 3.0334E-06 \\    
$\rm^{73}Ge $ &  2.0591E-07 & 2.3031E-07 & 3.9178E-07 & 5.0022E-07 & 5.9293E-07 & 8.7946E-07 \\    
$\rm^{74}Ge $ &  9.5125E-07 & 1.1044E-06 & 1.7488E-06 & 2.1565E-06 & 2.8528E-06 & 3.3420E-06 \\    
$\rm^{75}Ge $ &  4.9825E-10 & 2.2758E-10 & 3.4557E-10 & 8.2160E-10 & 1.6328E-09 & 6.9154E-09 \\    
$\rm^{76}Ge $ &  1.6990E-07 & 1.6950E-07 & 2.0147E-07 & 2.4299E-07 & 2.9246E-07 & 3.5669E-07 \\    
$\rm^{77}Ge $ &  3.8909E-10 & 3.2139E-11 & 2.7430E-11 & 1.1217E-10 & 2.1718E-10 & 6.0987E-10 \\    
$\rm^{71}As $ &  7.6601E-11 & 4.0773E-11 & 4.2299E-11 & 6.0023E-11 & 4.5362E-11 & 1.1452E-10 \\    
$\rm^{72}As $ &  1.1458E-10 & 6.4727E-11 & 4.8789E-11 & 8.6594E-11 & 6.9550E-11 & 1.9255E-10 \\    
$\rm^{73}As $ &  6.1986E-09 & 6.2215E-09 & 5.4479E-09 & 8.5075E-09 & 8.7931E-09 & 1.6844E-08 \\    
$\rm^{74}As $ &  1.5251E-09 & 1.0291E-09 & 1.1269E-09 & 1.6911E-09 & 1.6102E-09 & 3.2882E-09 \\    
$\rm^{75}As $ &  1.5189E-07 & 1.7946E-07 & 2.8603E-07 & 3.9806E-07 & 4.5631E-07 & 6.7481E-07 \\    
$\rm^{76}As $ &  4.8010E-09 & 2.8692E-09 & 5.8324E-09 & 1.0688E-08 & 1.4521E-08 & 1.4107E-08 \\    
$\rm^{77}As $ &  2.3001E-08 & 1.9538E-08 & 2.3370E-08 & 5.9587E-08 & 5.2682E-08 & 1.2877E-07 \\    
$\rm^{74}Se $ &  2.1809E-08 & 5.1847E-08 & 3.7281E-08 & 6.1599E-08 & 6.3026E-08 & 1.0163E-07 \\    
$\rm^{75}Se $ &  2.0656E-09 & 3.2780E-09 & 1.8457E-09 & 3.6055E-09 & 3.4456E-09 & 6.7704E-09 \\    
$\rm^{76}Se $ &  1.6429E-07 & 2.6150E-07 & 3.1912E-07 & 3.9498E-07 & 4.2730E-07 & 5.7994E-07 \\    
$\rm^{77}Se $ &  9.5247E-08 & 1.0709E-07 & 1.6280E-07 & 1.8107E-07 & 2.2259E-07 & 2.5024E-07 \\    
$\rm^{78}Se $ &  3.3849E-07 & 4.1270E-07 & 6.6193E-07 & 7.5859E-07 & 8.8446E-07 & 1.0479E-06 \\    
$\rm^{79}Se $ &  1.1500E-08 & 2.4539E-08 & 8.3149E-08 & 1.2033E-07 & 1.4032E-07 & 1.8917E-07 \\    
$\rm^{80}Se $ &  5.8077E-07 & 6.6736E-07 & 9.1681E-07 & 1.0614E-06 & 1.2538E-06 & 1.2772E-06 \\    
$\rm^{81}Se $ &  4.8098E-15 & 3.5457E-14 & 1.0367E-13 & 1.6082E-13 & 2.6248E-12 & 4.0604E-13 \\    
$\rm^{82}Se $ &  1.0252E-07 & 1.0390E-07 & 1.1686E-07 & 1.3876E-07 & 1.7319E-07 & 1.9847E-07 \\    
$\rm^{83}Se $ &  2.2870E-16 & 6.7307E-17 & 4.0455E-16 & 8.3141E-16 & 2.4316E-14 & 4.7419E-15 \\    
$\rm^{75}Br $ &  2.6788E-14 & 2.1969E-14 & 4.3990E-15 & 9.4556E-15 & 7.2679E-15 & 1.5473E-14 \\    
$\rm^{76}Br $ &  3.2990E-12 & 2.5117E-12 & 7.8393E-13 & 2.1535E-12 & 1.5155E-12 & 4.2665E-12 \\    
$\rm^{77}Br $ &  2.1545E-10 & 2.3986E-10 & 8.2007E-11 & 1.6664E-10 & 1.3115E-10 & 2.9818E-10 \\    
$\rm^{78}Br $ &  3.2577E-18 & 3.6296E-16 & 4.6719E-18 & 8.8279E-18 & 4.8390E-16 & 4.2895E-16 \\    
$\rm^{79}Br $ &  8.8304E-08 & 9.4990E-08 & 1.0742E-07 & 1.0974E-07 & 1.2075E-07 & 1.4083E-07 \\    
$\rm^{80}Br $ &  1.1617E-14 & 1.3936E-13 & 5.6969E-14 & 1.3560E-14 & 1.8812E-12 & 3.2670E-14 \\    
$\rm^{81}Br $ &  1.0101E-07 & 1.1670E-07 & 1.6399E-07 & 1.9846E-07 & 2.2821E-07 & 2.4614E-07 \\    
$\rm^{82}Br $ &  1.2652E-09 & 7.3967E-10 & 2.6800E-09 & 6.8734E-09 & 5.4880E-09 & 7.5944E-09 \\    
$\rm^{83}Br $ &  1.2726E-09 & 1.1242E-09 & 2.3666E-09 & 3.6737E-09 & 3.8932E-09 & 8.5207E-09 \\    
$\rm^{78}Kr $ &  3.5773E-09 & 5.8862E-09 & 6.1713E-09 & 7.5132E-09 & 7.9248E-09 & 8.8188E-09 \\    
$\rm^{79}Kr $ &  5.2914E-11 & 1.5469E-10 & 1.7288E-10 & 2.0514E-10 & 2.0198E-10 & 2.4355E-10 \\    
$\rm^{80}Kr $ &  3.1112E-08 & 4.3875E-08 & 5.5573E-08 & 6.2319E-08 & 6.5856E-08 & 8.3729E-08 \\    
$\rm^{81}Kr $ &  2.2405E-09 & 1.7787E-09 & 2.1518E-09 & 3.3229E-09 & 2.5094E-09 & 1.1228E-08 \\    
$\rm^{82}Kr $ &  1.3275E-07 & 1.5977E-07 & 2.1477E-07 & 2.4332E-07 & 2.5971E-07 & 3.2128E-07 \\    
$\rm^{83}Kr $ &  1.1741E-07 & 1.2765E-07 & 1.8704E-07 & 2.2328E-07 & 2.6196E-07 & 2.9958E-07 \\    
$\rm^{84}Kr $ &  5.3822E-07 & 6.0391E-07 & 8.6853E-07 & 9.5145E-07 & 1.0546E-06 & 1.3044E-06 \\    
$\rm^{85}Kr $ &  3.8801E-09 & 1.0503E-08 & 7.8585E-08 & 9.6109E-08 & 1.1096E-07 & 1.4629E-07 \\    
$\rm^{86}Kr $ &  1.7514E-07 & 2.0678E-07 & 3.4960E-07 & 3.6573E-07 & 4.5272E-07 & 4.6276E-07 \\    
$\rm^{87}Kr $ &  4.3110E-13 & 1.8238E-13 & 2.3086E-13 & 3.5864E-13 & 4.4656E-12 & 1.0655E-12 \\    
$\rm^{79}Rb $ &  1.7717E-19 & 5.4159E-18 & 3.2918E-19 & 7.0332E-19 & 3.4077E-18 & 2.4238E-18 \\    
$\rm^{80}Rb $ &  1.6475E-29 & 2.1507E-23 & 2.0306E-24 & 1.0395E-23 & 3.4209E-20 & 1.1507E-20 \\    
$\rm^{81}Rb $ &  1.6952E-13 & 4.7065E-13 & 1.0203E-13 & 1.8056E-13 & 1.6578E-13 & 2.4656E-13 \\    
$\rm^{82}Rb $ &  1.0570E-23 & 1.0164E-19 & 2.7588E-21 & 2.3190E-20 & 2.3877E-17 & 2.2047E-17 \\    
$\rm^{83}Rb $ &  5.9915E-11 & 2.0159E-10 & 2.5154E-10 & 2.8027E-10 & 3.0261E-10 & 3.3766E-10 \\    
$\rm^{84}Rb $ &  1.0992E-11 & 2.4247E-11 & 3.1103E-11 & 3.2857E-11 & 3.4713E-11 & 3.9312E-11 \\    
$\rm^{85}Rb $ &  1.2540E-07 & 1.4140E-07 & 1.7677E-07 & 1.9037E-07 & 2.0825E-07 & 2.4056E-07 \\    
$\rm^{86}Rb $ &  1.0118E-09 & 1.5671E-09 & 8.7724E-09 & 1.6323E-08 & 1.6782E-08 & 2.6928E-08 \\    
$\rm^{87}Rb $ &  5.9080E-08 & 7.5342E-08 & 1.6508E-07 & 1.8268E-07 & 2.6411E-07 & 2.2816E-07 \\    
$\rm^{88}Rb $ &  8.3467E-17 & 7.7150E-16 & 1.6388E-15 & 1.6489E-15 & 1.0371E-13 & 2.4812E-15 \\    
$\rm^{84}Sr $ &  2.5367E-09 & 6.4730E-09 & 5.6308E-09 & 7.9518E-09 & 8.9759E-09 & 9.1477E-09 \\    
$\rm^{85}Sr $ &  3.7021E-11 & 1.9585E-10 & 3.2083E-10 & 4.1970E-10 & 4.4458E-10 & 4.3649E-10 \\    
$\rm^{86}Sr $ &  5.1850E-08 & 6.2316E-08 & 9.7237E-08 & 1.0481E-07 & 1.0904E-07 & 1.5727E-07 \\    
$\rm^{87}Sr $ &  3.2313E-08 & 3.6787E-08 & 5.9854E-08 & 7.0472E-08 & 6.9504E-08 & 1.1785E-07 \\    
$\rm^{88}Sr $ &  3.4039E-07 & 4.0484E-07 & 7.2238E-07 & 7.8819E-07 & 8.8175E-07 & 1.0635E-06 \\    
$\rm^{89}Sr $ &  4.1677E-10 & 7.0856E-10 & 4.9472E-09 & 1.1788E-08 & 1.0276E-08 & 2.0529E-08 \\    
$\rm^{90}Sr $ &  1.8052E-10 & 2.9079E-10 & 3.2613E-09 & 4.3414E-09 & 5.3861E-09 & 4.6955E-09 \\    
$\rm^{91}Sr $ &  1.0210E-12 & 3.5163E-13 & 6.8911E-13 & 1.7310E-12 & 1.3232E-11 & 4.5358E-12 \\    
$\rm^{85}Y  $ &  3.5900E-16 & 5.0174E-15 & 4.7830E-16 & 1.2009E-15 & 2.0477E-15 & 1.8683E-15 \\    
$\rm^{86}Y  $ &  3.5406E-15 & 2.2803E-14 & 7.4563E-15 & 1.3710E-14 & 1.5624E-14 & 1.6705E-14 \\    
$\rm^{87}Y  $ &  1.2986E-12 & 9.2453E-12 & 6.5571E-12 & 9.3070E-12 & 9.9648E-12 & 1.0756E-11 \\    
$\rm^{88}Y  $ &  2.6218E-12 & 1.5272E-11 & 2.0019E-11 & 2.5780E-11 & 2.4941E-11 & 2.8607E-11 \\    
$\rm^{89}Y  $ &  8.7988E-08 & 1.0395E-07 & 1.4308E-07 & 1.5927E-07 & 1.6321E-07 & 2.0697E-07 \\    
$\rm^{90}Y  $ &  6.9429E-11 & 1.5954E-10 & 4.5396E-10 & 1.7080E-09 & 4.9852E-10 & 2.0204E-09 \\    
$\rm^{91}Y  $ &  2.3918E-10 & 4.3546E-10 & 5.5460E-09 & 6.3491E-09 & 7.3615E-09 & 7.3127E-09 \\    
$\rm^{90}Zr $ &  1.0455E-07 & 1.2110E-07 & 1.5304E-07 & 1.6628E-07 & 1.7762E-07 & 2.1338E-07 \\    
$\rm^{91}Zr $ &  2.3043E-08 & 2.5665E-08 & 3.5048E-08 & 3.6726E-08 & 3.9811E-08 & 4.8511E-08 \\    
$\rm^{92}Zr $ &  3.5736E-08 & 4.0114E-08 & 5.3434E-08 & 5.7498E-08 & 6.4215E-08 & 7.3767E-08 \\    
$\rm^{93}Zr $ &  4.1076E-10 & 7.1280E-10 & 2.8758E-09 & 3.7473E-09 & 3.7800E-09 & 6.8914E-09 \\    
$\rm^{94}Zr $ &  3.6514E-08 & 4.0178E-08 & 4.9844E-08 & 5.3795E-08 & 5.9805E-08 & 6.8377E-08 \\    
$\rm^{95}Zr $ &  5.2319E-11 & 7.2603E-11 & 2.9861E-10 & 6.7747E-10 & 8.9675E-10 & 1.8563E-09 \\    
$\rm^{96}Zr $ &  5.9762E-09 & 6.5773E-09 & 8.1290E-09 & 9.1949E-09 & 1.0282E-08 & 1.1270E-08 \\    
$\rm^{97}Zr $ &  1.3925E-13 & 3.0342E-14 & 2.3918E-14 & 7.6886E-14 & 4.7061E-12 & 1.7184E-13 \\    
$\rm^{91}Nb $ &  3.2627E-12 & 2.1556E-11 & 3.5595E-11 & 4.2141E-11 & 4.1385E-11 & 5.1969E-11 \\    
$\rm^{92}Nb $ &  8.4675E-13 & 3.8456E-12 & 5.8068E-12 & 6.6832E-12 & 5.9916E-12 & 7.8858E-12 \\    
$\rm^{93}Nb $ &  1.5159E-08 & 1.6459E-08 & 1.7938E-08 & 1.9389E-08 & 2.0728E-08 & 2.3292E-08 \\    
$\rm^{94}Nb $ &  3.0990E-11 & 1.3897E-11 & 1.8254E-11 & 3.3269E-11 & 4.9903E-11 & 1.1272E-10 \\    
$\rm^{95}Nb $ &  1.0160E-11 & 3.5069E-11 & 1.0813E-10 & 7.3676E-11 & 1.9877E-10 & 2.3045E-10 \\    
$\rm^{96}Nb $ &  1.3137E-12 & 2.7659E-12 & 2.7858E-12 & 3.5656E-12 & 7.2186E-12 & 1.2364E-11 \\    
$\rm^{97}Nb $ &  8.9272E-14 & 2.3152E-13 & 1.2831E-13 & 1.3184E-13 & 6.8219E-13 & 2.2064E-12 \\    
$\rm^{92}Mo $ &  5.8641E-09 & 6.6247E-09 & 7.7347E-09 & 8.4701E-09 & 9.2844E-09 & 1.0715E-08 \\    
$\rm^{93}Mo $ &  7.9438E-12 & 2.1147E-11 & 4.7725E-11 & 5.6314E-11 & 6.4561E-11 & 8.9948E-11 \\    
$\rm^{94}Mo $ &  4.0927E-09 & 4.6330E-09 & 5.3821E-09 & 5.9181E-09 & 6.4171E-09 & 7.4351E-09 \\    
$\rm^{95}Mo $ &  6.6351E-09 & 7.2360E-09 & 7.9775E-09 & 8.6798E-09 & 9.1654E-09 & 1.0438E-08 \\    
$\rm^{96}Mo $ &  7.2324E-09 & 7.9942E-09 & 9.1915E-09 & 1.0069E-08 & 1.0688E-08 & 1.2857E-08 \\    
$\rm^{97}Mo $ &  4.1015E-09 & 4.4793E-09 & 4.9587E-09 & 5.4757E-09 & 5.7804E-09 & 6.9226E-09 \\    
$\rm^{98}Mo $ &  1.0631E-08 & 1.1627E-08 & 1.3089E-08 & 1.4352E-08 & 1.5252E-08 & 1.8405E-08 \\    
$\rm^{132}Xe$ &  3.5335E-08 & 3.8605E-08 & 4.4443E-08 & 4.8085E-08 & 5.2145E-08 & 6.0427E-08 \\    
$\rm^{133}Xe$ &  1.4856E-11 & 1.4693E-11 & 2.5237E-11 & 1.7896E-10 & 8.0201E-11 & 3.9454E-10 \\    
$\rm^{134}Xe$ &  1.3209E-08 & 1.4366E-08 & 1.6184E-08 & 1.7687E-08 & 1.9452E-08 & 2.1838E-08 \\    
$\rm^{135}Xe$ &  2.5247E-12 & 1.4336E-12 & 1.5307E-12 & 4.4271E-12 & 5.0759E-12 & 2.5235E-11 \\    
$\rm^{133}Cs$ &  9.0974E-09 & 9.9211E-09 & 1.0982E-08 & 1.1834E-08 & 1.2871E-08 & 1.4732E-08 \\    
$\rm^{134}Cs$ &  7.5308E-12 & 1.7218E-11 & 4.0461E-11 & 7.3428E-11 & 1.0541E-10 & 2.5222E-10 \\    
$\rm^{135}Cs$ &  7.6729E-11 & 1.6275E-10 & 5.3369E-10 & 7.7141E-10 & 8.7549E-10 & 1.5379E-09 \\    
$\rm^{136}Cs$ &  3.9485E-12 & 7.0735E-12 & 2.2252E-11 & 1.4480E-10 & 9.3027E-11 & 3.6279E-10 \\    
$\rm^{137}Cs$ &  2.3759E-11 & 8.5386E-11 & 6.7744E-10 & 9.9224E-10 & 1.3121E-09 & 1.4036E-09 \\    
$\rm^{138}Cs$ &  5.1614E-19 & 3.8889E-18 & 2.4028E-18 & 3.0245E-17 & 1.1358E-15 & 1.1058E-17 \\    
$\rm^{134}Ba$ &  3.1984E-09 & 3.5205E-09 & 3.9953E-09 & 4.3980E-09 & 4.8241E-09 & 5.7403E-09 \\    
$\rm^{135}Ba$ &  7.5900E-09 & 8.2202E-09 & 8.9519E-09 & 9.6741E-09 & 1.0350E-08 & 1.1707E-08 \\    
$\rm^{136}Ba$ &  9.7697E-09 & 1.0789E-08 & 1.2474E-08 & 1.3735E-08 & 1.5117E-08 & 1.8720E-08 \\    
$\rm^{137}Ba$ &  1.3569E-08 & 1.4810E-08 & 1.6437E-08 & 1.8099E-08 & 1.9371E-08 & 2.3649E-08 \\    
$\rm^{138}Ba$ &  8.7772E-08 & 9.6129E-08 & 1.1016E-07 & 1.2250E-07 & 1.3518E-07 & 1.6559E-07 \\    
$\rm^{139}Ba$ &  1.1701E-14 & 6.0165E-15 & 1.3539E-15 & 1.1166E-14 & 3.9433E-13 & 5.5033E-15 \\    
$\rm^{138}La$ &  8.8564E-12 & 9.6625E-12 & 1.0662E-11 & 1.1565E-11 & 1.2507E-11 & 1.4105E-11 \\    
$\rm^{139}La$ &  1.0227E-08 & 1.1200E-08 & 1.2706E-08 & 1.4158E-08 & 1.5663E-08 & 1.9093E-08 \\    
$\rm^{140}La$ &  3.5018E-13 & 7.4634E-14 & 1.6542E-14 & 1.2587E-13 & 7.3539E-12 & 8.6683E-14 \\    
$\rm^{140}Ce$ &  2.7444E-08 & 2.9929E-08 & 3.3451E-08 & 3.6912E-08 & 4.0361E-08 & 4.7756E-08 \\    
$\rm^{141}Ce$ &  2.2223E-12 & 5.9355E-13 & 4.2013E-13 & 1.4019E-12 & 7.5892E-11 & 1.0293E-11 \\    
$\rm^{141}Pr$ &  4.2707E-09 & 4.6512E-09 & 5.1944E-09 & 5.7174E-09 & 6.0957E-09 & 7.0322E-09 \\    
$\rm^{142}Pr$ &  5.5365E-13 & 9.0732E-14 & 4.9165E-14 & 2.2092E-13 & 4.8494E-12 & 4.3948E-13 \\    
$\rm^{142}Nd$ &  5.9629E-09 & 6.5907E-09 & 7.5667E-09 & 8.4529E-09 & 9.5033E-09 & 1.1348E-08 \\    
$\rm^{143}Nd$ &  2.5719E-09 & 2.7913E-09 & 3.0445E-09 & 3.3028E-09 & 3.5479E-09 & 4.0634E-09 \\    
$\rm^{144}Nd$ &  5.2375E-09 & 5.6941E-09 & 6.2646E-09 & 6.8423E-09 & 7.4261E-09 & 8.5919E-09 \\    
$\rm^{202}Hg$ &  5.2295E-09 & 5.6921E-09 & 6.3973E-09 & 7.1127E-09 & 7.6773E-09 & 9.3406E-09 \\    
$\rm^{203}Hg$ &  3.4365E-12 & 2.3568E-12 & 1.1833E-12 & 5.7036E-12 & 6.8187E-11 & 1.2047E-10 \\    
$\rm^{204}Hg$ &  1.1964E-09 & 1.3029E-09 & 1.4324E-09 & 1.5858E-09 & 1.7571E-09 & 1.9617E-09 \\    
$\rm^{205}Hg$ &  4.6532E-21 & 5.7901E-20 & 2.6851E-20 & 5.5695E-21 & 1.2002E-18 & 6.1592E-22 \\    
$\rm^{203}Tl$ &  2.7602E-09 & 3.0077E-09 & 3.3520E-09 & 3.6814E-09 & 3.9389E-09 & 4.6783E-09 \\    
$\rm^{204}Tl$ &  2.7128E-12 & 5.3998E-12 & 2.2216E-11 & 4.5463E-11 & 4.0681E-11 & 1.5096E-10 \\    
$\rm^{205}Tl$ &  6.6272E-09 & 7.2086E-09 & 7.9334E-09 & 8.6921E-09 & 9.3051E-09 & 1.0782E-08 \\    
$\rm^{206}Tl$ &  4.9024E-20 & 1.2224E-19 & 8.2441E-20 & 7.1406E-20 & 2.4017E-17 & 7.1165E-22 \\    
$\rm^{204}Pb$ &  1.3691E-09 & 1.5031E-09 & 1.7027E-09 & 1.8700E-09 & 2.1892E-09 & 2.5191E-09 \\    
$\rm^{205}Pb$ &  2.3542E-11 & 2.8221E-11 & 4.8578E-11 & 7.1702E-11 & 1.3344E-10 & 2.8824E-10 \\    
$\rm^{206}Pb$ &  1.2578E-08 & 1.3737E-08 & 1.5680E-08 & 1.7309E-08 & 1.9561E-08 & 2.3007E-08 \\    
$\rm^{207}Pb$ &  1.3124E-08 & 1.4187E-08 & 1.5425E-08 & 1.6967E-08 & 1.8124E-08 & 2.1982E-08 \\    
$\rm^{208}Pb$ &  3.7676E-08 & 4.0691E-08 & 4.4559E-08 & 4.8743E-08 & 5.2347E-08 & 6.1776E-08 \\    
$\rm^{209}Pb$ &  2.1274E-16 & 2.1841E-16 & 2.4099E-16 & 1.9698E-15 & 1.1240E-13 & 7.7855E-16 \\    
$\rm^{208}Bi$ &  2.5878E-11 & 5.5137E-11 & 1.1631E-10 & 1.5225E-10 & 2.6972E-10 & 2.9627E-10 \\    
$\rm^{209}Bi$ &  5.0879E-09 & 5.4854E-09 & 5.9526E-09 & 6.4658E-09 & 6.8909E-09 & 7.9884E-09 \\    
\enddata                  
\end{deluxetable}

\begin{deluxetable*}{lc}\label{tab_expl_prop}
\tablecolumns{2}
\tablecaption{Initial mass-explosion energy relation adopted for the light curves calcutions}
\tablehead{\colhead{Mass ($\rm M_\odot$)} & \colhead{$E_{\rm expl}$ ($\rm 10^{50}~erg$)} } 
\startdata    
9.22 & 1.24 \\
10   & 2.13 \\
11   & 3.26 \\
12   & 3.33 \\
13   & 3.39 \\
15   & 3.52 \\
\enddata
\end{deluxetable*}

\begin{acknowledgments}
This work has been mainly supported by the Theory Grant "Evolution, nucleosynthesis and final fate of stars in the transition between AGB and Massive Stars" (PI M. Limongi) of the INAF Fundamental Astrophysics Funding Program 2022-2023. This work has also been partially supported by the World Premier International Research Center Initiative (WPI), MEXT, Japan. KN has been supported by the Japan Society for the Promotion of Science (JSPS) KAKENHI grants JP20K04024, JP21H04499, and JP23K03452. 
\end{acknowledgments}

\end{document}